# A Discrete Macro-Element Method (DMEM) for the nonlinear structural assessment of masonry arches

F. Cannizzaro[1], B. Pantò[1], S. Caddemi[1], I. Caliò[1]

Department of Civil Engineering and Architecture, University of Catania, Italy

**Abstract**

The structural response of masonry arches is strongly dominated by the arch geometry, the stone block dimensions and the interaction with backfill material or surrounding walls. Due to their intrinsic discontinuous nature, the nonlinear structural response of these key historical structures can be efficiently modelled in the context of discrete element approaches. Smeared crack finite elements models, based on the assumption of homogenised media and spread plasticity, fail to rigorously predict the actual collapse behaviour of such structures, that are generally governed by rocking and sliding mechanisms along mortar joints between stone blocks. In this paper a new Discrete Macro-Element Method (DMEM) for predicting the nonlinear structural behaviour of masonry arches is proposed. The method is based on a macro-element discretization in which each plane element interacts with the adjacent elements through zero-thickness interfaces and whose internal deformability is related to a single degree of freedom only. Both experimental and numerical validations show the capability of the proposed approach to be applied for the prediction of the non-linear response of masonry arch structures under different loading conditions.

## 1. Introduction

Although arches, vaults and domes have been adopted since ancient ages [1] for engineering works, their complete structural assessment is not an easy task even today. Masonry arches transmit the self-weight and the applied loads through load-paths that mainly involve compressive stresses by taking advantage of gravity loads through their own shape. The high nonlinearity, due to low-tensile resistance of masonry or to the presence of dry stone-interfaces, does not allow the assumption of linear elastic behaviour and leads to load dependent equilibrium solutions strongly related to the arch geometry and its supports conditions. A further very complex numerical issue is related to the presence of the backfill whose actual structural contribution is very difficult to model due to the non cohesive nature of the material generally adopted [2]. For this reason in many cases it is generally preferred to model the backfill structural role simply considering its stabilising effect related to its own weight and neglecting its mechanical contribution.

In the past, graphical based design approaches have been developed and widely used for the structural design and the construction of monumental structures [3]. However, these traditional methods, based on the concept of the line of thrust, are difficult to apply in presence of material nonlinearities and cyclic loads related to dynamic actions such as earthquakes. On the other hand, the potential availability of efficient and easy to apply numerical approaches could allow performing nonlinear structural analyses under different loading conditions, such as dynamic excitations or moving loads, these latter typical of masonry arch bridges.

The most important contributions to the understanding of the structural behaviour of stone and masonry arches were provided by Jacques Heyman in his famous treatises [3]-[7]. More recently, several numerical strategies have been proposed based on linear [8]-[11] or nonlinear Finite Element Models (FEM) [12],[13], or Discrete Element Method (DEM) [14],[15].

Sarhosis et al. [16] presented a three dimensional computational model, based on the DEM, which was used to investigate the effect of the angle of skew on the load carrying capacity of twenty-eight single span stone masonry arches with different geometric layouts.

Rizzi et al. [17] presented an analytical and numerical analysis of the classical Couplet–Heyman problem in the statics of circular masonry arches.

Dimitri and Tornabene [18] developed an analytical model based on limit analysis for describing the stability of pointed and basket-handle arches and portals with respect to circular ones, for varying geometry parameters. They compared the predictions of the analytical model with results of numerical modelling by the classical DEM and obtained a satisfactory agreement showing the potentiality of the discrete element framework as a method of evaluating the quasi-static behaviour of unreinforced masonry structures. In the context of the DEM strategies, Dimitri et al. [19] and De Lorenzis et al. [20],[21] investigated the important role of buttresses, in the dynamic field, considering several shapes of the buttress, typical of ancient constructions.

Gago et al. in [2], using modern structural analysis, explained the favourable effect of the extrados infill in the stability of arched structures also highlighting the high collapse risk related to the backfill removal.

Very recently Zhang et al [22] investigated the nonlinear response of brick-masonry arches, up to collapse, by using an accurate 3D meso-scale description utilising solid elements for representing brick units and 2D nonlinear interface elements for describing mortar joints and brick-mortar interfaces. The masonry meso-scale strategy has been also combined with an original domain partitioning approach that, allowing for parallel computation, leads to powerful high accurate computational tool applicable for large structures.

In this paper an innovative Discrete Macro-Element Method (DMEM), alternative to previously proposed approaches, for the simulation of the nonlinear behaviour of masonry arches is presented. The proposed approach takes advantage of the Discrete Element Method (DEM) strategies at a 'macro-scale'. Differently from the classical DEM approach, in which each element is considered as a rigid body, in the proposed DMEM strategy each macro-element possesses a shear deformability allowing to identify shear diagonal local failure. This shear deformability is related to a single degree of freedom for each macro-element. The mechanical interaction among adjacent macro-elements is concentrated in zero-thickness interfaces distributed along the entire length of the contact edges. The computational cost of the proposed numerical approach is greatly reduced in comparison to that involved in detailed nonlinear finite element simulations or DEM strategies based on meso-scale discretizations.

The basic macro-element, which is adopted for the simulation of an arch macro-portion, is described in Section 2. It consists of an articulated irregular quadrilateral whose internal deformability is dependent on a single degree of freedom. Three further Lagrangian parameters identify the rigid motion of the element needed to complete its plane kinematics description. The flexural and shear-sliding behaviours are governed by along-edge zero-thickness interfaces, lying on the sides of the quadrilateral and governing the interaction with the adjacent macro-elements. Three specific non-linear one-dimensional constitutive laws are considered in the model for simulating separately the flexural, shear-diagonal and shear-sliding behaviour of the masonry medium, assumed as an orthotropic homogenised continuous material. The calibration of the model requires few parameters in order to define the basic masonry mechanical properties. Such properties of the material can be easily obtained from current experimental tests and/or suggested by technical codes. The equivalence between the masonry arch portion, that is represented at the macro-scale, and the macro-element is here based on a very simple fibre calibration approach making the interpretation of the numerical results straightforward and unambiguous.

In Section 2 a detailed description of the kinematics of the proposed macro-element is provided. Furthermore, a qualitative classification of the typical failure mechanism of masonry arches is presented and it is shown how the proposed approach is capable to provide a satisfactory prediction of all the possible collapse mechanisms of a masonry arch.

In Section 3 the mechanical calibration of the element in terms of its shear behaviour and its contouring interfaces, which govern the membrane deformability of the element, are described. In the numerical applications (Section 4), the model is applied for the simulation of the nonlinear response of masonry arches for which experimental and numerical results are available from previous research-studies reported in the literature.

The obtained results show the capability of the proposed 'parsimonious' (i.e. low cost) approach to be used for the structural assessment of masonry arch structures both for researches and practical applications.

## 2. The DMEM formulation for masonry arches

The proposed nonlinear discrete macro-element for plane masonry curved structural elements, such as arches, is defined according to an original approach that enriches the classical discrete element strategy generally based on rigid elements interacting by means of nonlinear nonlinear links.

The basic element here proposed can be described through a mechanical representation obtained as a nontrivial upgrade of a regular macro-element, based on the use of rectangular elements, conceived for the simulation of unreinforced and confined masonry structures [23]-[26]. In this new formulation the element is conceived as a plane irregular articulated quadrilateral, formed by four rigid sides connected by internal hinges, hence, differently from the classical DEM, the element is endowed with an internal shear deformability that is related to a single degree of freedom. Furthermore, the interaction between adjacent elements is modelled by along-sides nonlinear zero-thickness continuous interfaces, Figure 1. The latter, connecting two rigid sides of different quadrilaterals, are responsible for the axial and flexural behaviour as well as the shear sliding between adjacent elements.

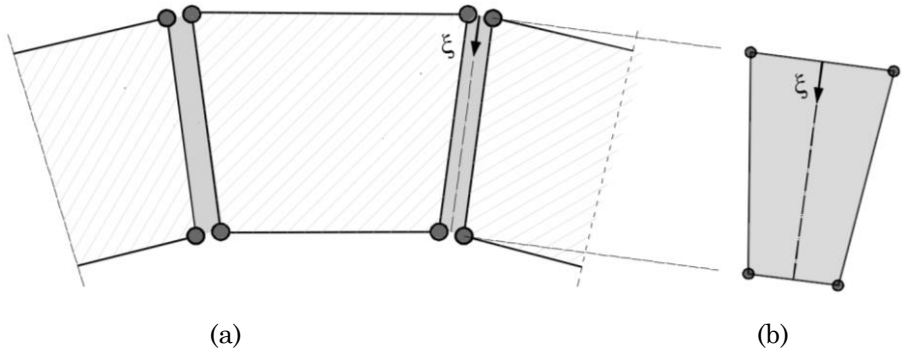

(a) (b)

Figure 1. The discrete macro-element: (a) discretization pattern of a masonry arch; (b) the interface for the case of a symmetric variable cross section.

The kinematics of the proposed plane macro-element, although described by four degrees of freedom only, allows a simple but accurate description of the flexural, shear diagonal and shear sliding collapse behaviour of masonry arches. Thanks to the capability of capturing the main collapse mechanisms, the above introduced macro-element modelling leads to an efficient simulation of arch structures loaded in their own plane.

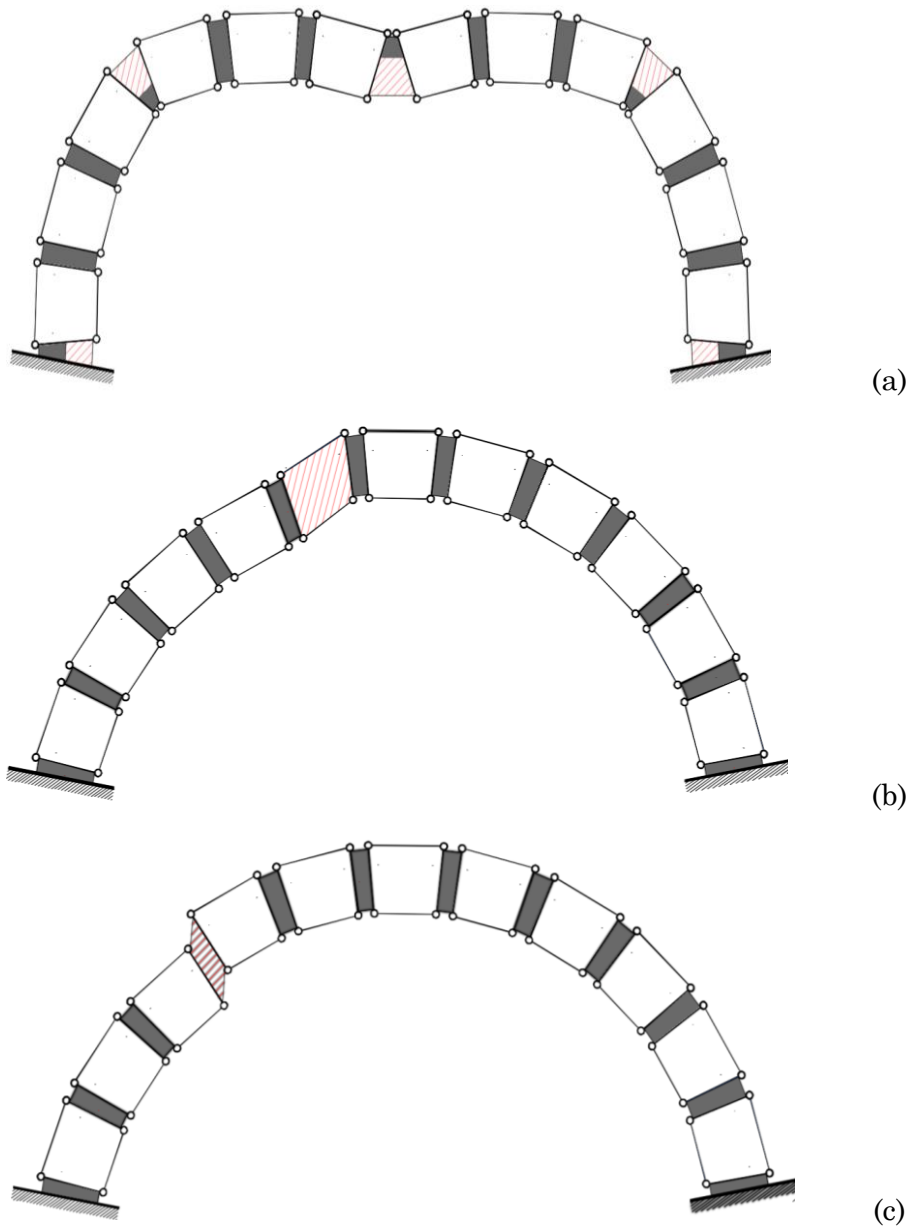

Figure 2. Typical in-plane collapse mechanisms of a masonry arch: (a) Flexural failure scenario related to the formation of several hinges; (b) Shear failure scenario due to the failure of a stone element or a finite portion of a masonry arch; (c) Shear failure scenario due to the localised sliding along a mortar joint.

Some typical arch collapse scenarios, in which the relevant damage patterns are highlighted, are reported in Figure 2. Namely, Figure 2a reports an arch collapse mechanism dominated by a flexural failure mechanism, in which five hinges are activated. Figure 2b shows the capability of the element to grasp shear local crisis of a finite portion of masonry element. Figure 2c highlights a shear failure scenario related to shear sliding mechanism between stone blocks or concentrated in mortar joints.

It is worth to notice that the overall elastic shear deformation of a masonry arch, discretized by several macro-elements, can be partly related to the diagonal deformations of the irregular quadrilateral and partly attributed to concentrated sliding displacements along the interfaces. In the proposed formulation, the shear-sliding mechanism is aimed to govern the deformations between macro-elements related to the occurrence of sliding along the interfaces (Fig.2c). On the other hand, the shear-type deformability enriches the element kinematics and, differently from the classical DEM approach, allows the simulation of possible diagonal shear cracking damage distribution or failure related to the shear collapse of a masonry element (Fig.2b). Mixed failure mechanisms can also be considered being each mechanism governed by specific constitutive laws.

The versatile geometry of the element allows a consistent simulation of masonry arch structures also in presence of complex geometrical layouts or for those cases in which the texture, related to the orientation of the mortar joints, could strongly affect the structural behaviour. In these latter cases, the proposed model can be employed considering the actual geometry and the real arrangement of the units through a consistent mesh of irregular quadrilateral elements. One peculiarity of the proposed approach is that the axial and flexural deformations of the homogenised masonry arch portion, represented by the macro-element, are both lumped in the zero-thickness interfaces.

## 2.1 Kinematics

The DME kinematics is presented in this section to describe both the relative displacements at the interface between elements and the shear deformation mechanism of the single macro-element.

The macro-element is constituted by an irregular plane quadrilateral whose edges, connected by four rotational hinges, are assumed to be rigid. Each side of the quadrilateral is characterized by its length $l_k$, $k=1,\ldots,4$, and the internal angles between the adjacent edges $\alpha_k$, $k=1,\ldots,4$, as indicated in Figure 3 (anti-clockwise numbering is adopted). The in-plane kinematics of each element is governed by 4 degrees of freedoms, three associated to the rigid body motion and the other related to the quadrilateral in-plane articulation. The chosen Lagrangian parameters, indicated in Figure 3, are the translational displacements $U$, $V$ and the rotation $\Phi$ of the centre of mass of the element '$G$' (Figure 3a) and the parameter $\Gamma$, that identifies the variation of the angle $\alpha_1$ between the edges departing from the origin of the local element reference system ($\mathbf{e}_x, \mathbf{e}_y$), as depicted in Figure 3b. All the Lagrangian parameters of the macro-element are collected in the vector $\mathbf{d}^T = \begin{bmatrix} U & V & \Phi & \Gamma \end{bmatrix}$.

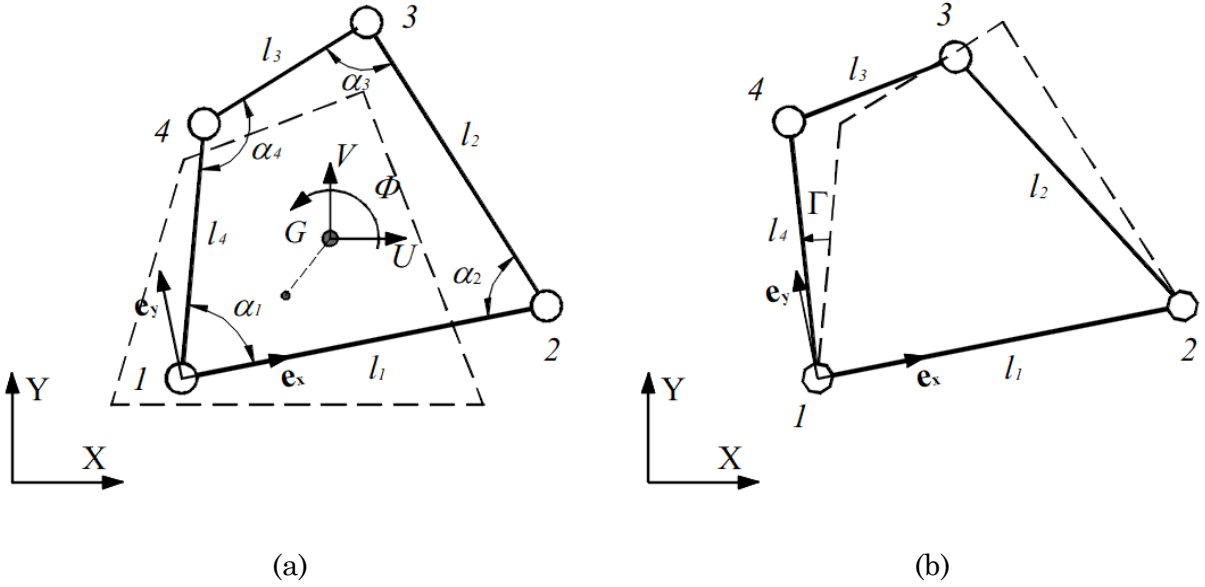

(a)            (b)

Figure 3. The element's kinematics and the chosen Lagrangian parameters: (a) the rigid body motion and (b) the generalized shear distortion

In order to describe the mechanical behaviour related to the interaction with adjacent elements, the definition of the in-plane kinematics of each side of the quadrilateral, as a function of the chosen Lagrangian parameters, is introduced in the following subparagraph.

### 2.1.1 Interface displacements

Let us consider two adjacent elements $p$ and $q$ sharing the $i$-th interface, where the $p$-th element is located on the left of the interface whereas the $q$-th element is located at its right, Figure 4. Denoting by $\xi$ the normalised abscissa, variable between 0 and 1, referred to a local reference system ($\mathbf{e}_\xi, \mathbf{e}_\eta$) of the $i$-th interface, the corresponding local longitudinal and orthogonal displacements of the two opposite element edges $u_p(\xi)$, $v_p(\xi)$ and $u_q(\xi)$, $v_q(\xi)$ can be expressed as function of corresponding auxiliary local degrees of freedom $u_p, v_{0p}, v_{1p}$ and $u_q, v_{0q}, v_{1q}$ (Figure 4) given by the displacements of the hinges at $\xi = 0, 1$, as follows:

$$\begin{bmatrix} u_p(\xi) \\ v_p(\xi) \end{bmatrix} = \begin{bmatrix} 1 & 0 & 0 \\ 0 & 1-\xi & \xi \end{bmatrix} \begin{bmatrix} u_p \\ v_{0p} \\ v_{1p} \end{bmatrix} \quad ; \quad \begin{bmatrix} u_q(\xi) \\ v_q(\xi) \end{bmatrix} = \begin{bmatrix} 1 & 0 & 0 \\ 0 & 1-\xi & \xi \end{bmatrix} \begin{bmatrix} u_q \\ v_{0q} \\ v_{1q} \end{bmatrix} \quad (1)$$

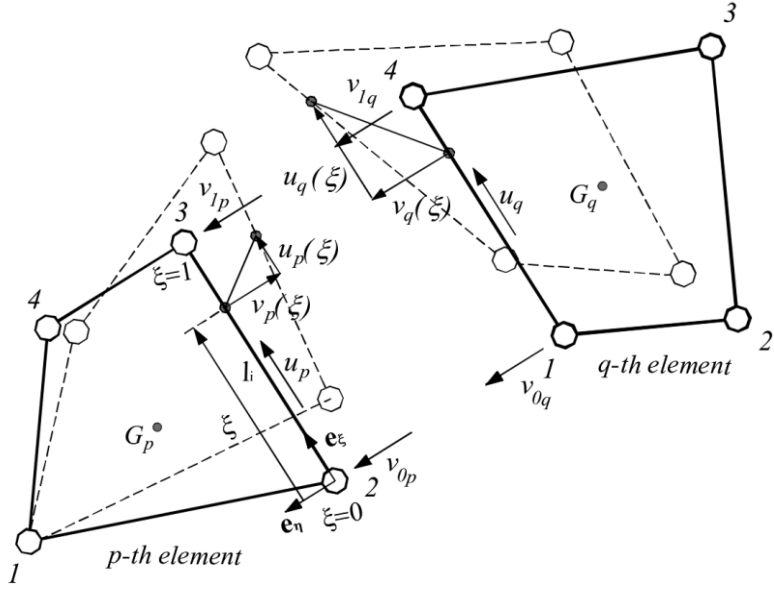

Figure 4. Local relative displacements in the interface between two adjacent macro-elements.

By collecting the local longitudinal and transversal displacement functions in the vectors $\mathbf{u}_p^T(\xi) = \begin{bmatrix} u_p(\xi) & v_p(\xi) \end{bmatrix}$ and $\mathbf{u}_q^T(\xi) = \begin{bmatrix} u_q(\xi) & v_q(\xi) \end{bmatrix}$, and the auxiliary local degrees of freedom of each edge in the vectors $\bar{\mathbf{u}}_p^T = \begin{bmatrix} u_p & v_{0p} & v_{1p} \end{bmatrix}$ and $\bar{\mathbf{u}}_q^T = \begin{bmatrix} u_q & v_{0q} & v_{1q} \end{bmatrix}$, Eqs.(1) can be rewritten in compact notation as follows:

$$\mathbf{u}_p(\xi) = \mathbf{N}(\xi)\bar{\mathbf{u}}_p \quad ; \quad \mathbf{u}_q(\xi) = \mathbf{N}(\xi)\bar{\mathbf{u}}_q \quad (2)$$

where

$$\mathbf{N}(\xi) = \begin{bmatrix} 1 & 0 & 0 \\ 0 & 1-\xi & \xi \end{bmatrix} \quad (3)$$

The auxiliary local degrees of freedom of each edge of the two adjacent elements $p,q$ at the $i$-th interface can be related to the relevant Lagrangian parameters as follows:

$$\bar{\mathbf{u}}_p = \mathbf{A}_p \mathbf{d}_p \qquad \bar{\mathbf{u}}_q = \mathbf{A}_q \mathbf{d}_q \quad (4)$$

where $\mathbf{A}_p, \mathbf{A}_q$ are compatibility matrices, whose components are simply related to the element geometry. The compatibility matrices, relative to the example reported in Figure 4, are given as follows:

$$\mathbf{A}_p = \begin{bmatrix} \mathbf{e}_\xi \cdot \mathbf{e}_{px} & \mathbf{e}_\xi \cdot \mathbf{e}_{py} & \left[-\left(y_{2p}-y_{Gp}\right)\mathbf{e}_{px}+\left(x_{2p}-x_{Gp}\right)\mathbf{e}_{py}\right]\cdot \mathbf{e}_\xi & 0 \\ \mathbf{e}_\eta \cdot \mathbf{e}_{px} & \mathbf{e}_\eta \cdot \mathbf{e}_{py} & \left[-\left(y_{2p}-y_{Gp}\right)\mathbf{e}_{px}+\left(x_{2p}-x_{Gp}\right)\mathbf{e}_{py}\right]\cdot \mathbf{e}_\eta & 0 \\ \mathbf{e}_\eta \cdot \mathbf{e}_{px} & \mathbf{e}_\eta \cdot \mathbf{e}_{py} & \left[-\left(y_{3p}-y_{Gp}\right)\mathbf{e}_{px}+\left(x_{3p}-x_{Gp}\right)\mathbf{e}_{py}\right]\cdot \mathbf{e}_\eta & -l_4\dfrac{\sin \alpha_4}{\sin \alpha_3} \end{bmatrix} \quad (5)$$

$$\mathbf{A}_q = \begin{bmatrix} \mathbf{e}_{1i} \cdot \mathbf{e}_{qx} & \mathbf{e}_{1i} \cdot \mathbf{e}_{qy} & \left[-\left(y_{1q}-y_{Gq}\right)\mathbf{e}_{qx}+\left(x_{1q}-x_{Gq}\right)\mathbf{e}_{qy}\right]\cdot \mathbf{e}_\xi & 0 \\ \mathbf{e}_{2i} \cdot \mathbf{e}_{qx} & \mathbf{e}_{2i} \cdot \mathbf{e}_{qy} & \left[-\left(y_{1q}-y_{Gq}\right)\mathbf{e}_{px}+\left(x_{1q}-x_{Gq}\right)\mathbf{e}_{qy}\right]\cdot \mathbf{e}_\eta & 0 \\ \mathbf{e}_{2i} \cdot \mathbf{e}_{qx} & \mathbf{e}_{2i} \cdot \mathbf{e}_{qy} & \left[-\left(y_{4q}-y_{Gq}\right)\mathbf{e}_{qx}+\left(x_{4q}-x_{Gq}\right)\mathbf{e}_{qy}\right]\cdot \mathbf{e}_\eta & l_4 \end{bmatrix}$$

where $\left(x_{Gp}, y_{Gp}\right)$ and $\left(x_{Gq}, y_{Gq}\right)$ represent the coordinates of the centre of mass of the *p*-th and *q*-th elements respectively. In view of Eq. (4), providing the displacements of the elements edges, Eq. (2), can be expressed as a function of the corresponding element degrees of freedoms as follows:

$$\mathbf{u}_p(\xi) = \mathbf{N}(\xi)\mathbf{A}_p \mathbf{d}_p \quad ; \quad \mathbf{u}_q(\xi) = \mathbf{N}(\xi)\mathbf{A}_q \mathbf{d}_q \quad (6)$$

therefore, the relative displacement function $\hat{\mathbf{u}}(\xi) = \mathbf{u}_q(\xi) - \mathbf{u}_p(\xi)$ of the *i*-th interface can be expressed as follows:

$$\hat{\mathbf{u}}(\xi) = \mathbf{N}(\xi)\mathbf{A}_q \mathbf{d}_q - \mathbf{N}(\xi)\mathbf{A}_p \mathbf{d}_p \quad (7)$$

### 2.1.2 The kinematics associated to the shear distortion $\Gamma$

Besides the kinematic characterisation of the interfaces between macro-elements, aiming at modelling the axial, bending and sliding behaviour of curved structures, the shear behaviour is intended to be described by the kinematics of the macro-element itself that implies angle variation of adjacent edges. The macro-element internal deformability, related to the Langrangian parameter $\Gamma$, implies displacements in correspondence of the 3rd and 4th vertexes of the element given by

$$u_{3x} = -\dfrac{l_4 \sin \alpha_2 \sin \alpha_4}{\sin \alpha_3}\Gamma; \quad u_{3y} = -\dfrac{l_4 \cos \alpha_2 \sin \alpha_4}{\sin \alpha_3}\Gamma; \quad u_{4x} = -l_4 \sin \alpha_1 \Gamma; \quad u_{4y} = l_4 \cos \alpha_1 \Gamma \quad (8)$$

## 3. The mechanical behaviour

The formulation here proposed follows a phenomenological description of the mechanical behaviour of an arch portion in which the zero-thickness interfaces rule the axial-flexural response and the shear sliding behaviour of adjacent elements, while the in-plane shear element deformability is related to the angular distortion of the articulated quadrilateral. The mechanical characterization of the zero-thickness interfaces is here performed through a straightforward fibre calibration procedure while the shear element deformability is calibrated through a mechanical equivalence with a reference geometric-consistent continuous plane model.

Each macro-element is intended to represent an equivalent homogenized masonry portion, whose mechanical properties can be inferred according to suitable homogenization techniques [27]-[29] here conveniently extended.

### *3.1. The interface stiffness matrix*

A peculiar aspect of the proposed numerical method is that the mechanical properties in the zero-thickness interfaces include both the stone and the mortar joints mechanical behaviour of the adjacent elements, leading to a simple homogenization strategy. As a consequence, the apparent interpenetration of the rigid edges of adjacent panels, along the zero-thickness interfaces, does not point out a compatibility violation but simply identifies states corresponding to compressive strains.

The zero-thickness continuous interfaces are characterised by a nonlinear behaviour described by the incremental relationship between the increments of the internal force distributions along the longitudinal and orthogonal directions of the interface, $df_u(\xi), df_v(\xi)$, collected in the vector $d\mathbf{f}_{int}^T(\xi) = \begin{bmatrix} df_u(\xi) & df_v(\xi) \end{bmatrix}$ and the relative displacement increment $d\hat{\mathbf{u}}(\xi)$ of the vector $\hat{\mathbf{u}}(\xi)$, introduced in Eq. (7), as follows:

$$d\mathbf{f}_{int}(\xi) = \mathbf{k}_T(\xi) d\hat{\mathbf{u}}(\xi) \tag{9}$$

where $\mathbf{k}_T(\xi)$ represents a $2 \times 2$ tangent stiffness distribution of the *i*-th interface at the abscissa $\xi$ that can be defined as follows:

$$\mathbf{k}_T(\xi) = \begin{bmatrix} k_{T_u}(\xi) & k_{T_{uv}}(\xi) \\ k_{T_{vu}}(\xi) & k_{T_v}(\xi) \end{bmatrix} \tag{10}$$

where the subscripts $u$ and $v$ identify the longitudinal and orthogonal directions of the generic $i$-th interface.

In view of Eq. (7), the force increment of the $i$-th interface $d\mathbf{f}_{int}(\xi)$, given by Eq. (9), can be also expressed as function of the degrees of freedom increments of the corresponding adjacent elements, denoted as $d\mathbf{d}_p, d\mathbf{d}_q$, as follows:

$$d\mathbf{f}_{int}(\xi) = \mathbf{k}_T(\xi)\left[\mathbf{N}(\xi)\mathbf{A}_q d\mathbf{d}_q - \mathbf{N}(\xi)\mathbf{A}_p d\mathbf{d}_p\right] \quad (11)$$

Considering Eq. (11) and by applying the principle of virtual work an 8×8 tangent stiffness matrix related to the contribution of the $i$-th interface, with respect to the global degrees of freedom of the two adjacent elements $p,q$, is obtained as follows:

$$\mathbf{K}_T = \int_0^1 \mathbf{A}^T \tilde{\mathbf{N}}^T(\xi)\mathbf{k}_T(\xi)\tilde{\mathbf{N}}(\xi)\mathbf{A}\,d\xi \quad (12)$$

being

$$\mathbf{A} = \begin{bmatrix} \mathbf{A}_q & \mathbf{0} \\ \mathbf{0} & \mathbf{A}_p \end{bmatrix}; \qquad \tilde{\mathbf{N}}(\xi) = \begin{bmatrix} \mathbf{N}(\xi) & -\mathbf{N}(\xi) \end{bmatrix} \quad (13)$$

The tangent stiffness matrix $\mathbf{K}_T$ rules the nonlinear behaviour of the $i$-th interface and its current value is related to the tangent interface stiffness distribution $\mathbf{k}_T(\xi)$.

The integration of Eq (12) has been performed according to a uniform fibre discretation of the adjacent elements, as depicted in Figure 5, where the masonry macro-elements $p$ and $q$, have been discretised according to $n_f$ cells of the homogenised masonry arch, and for simplicity, a constant thickness of the arch elements is considered.

Precisely, the contribution $\mathbf{k}_T(\xi_j)$ to the stiffness matrix of the $j$-th fibre at abscissa $\xi_j$, $j=1,\ldots,n_f$, is obtained by following a detailed calibration procedure based on the main mechanical and geometrical parameters of the masonry. According to the presented procedure, for the interface stiffness matrix definition, the generic interface is representative of the elastic/inelastic axial, flexural and sliding behaviours of adjacent finite portions of masonry considered as an equivalent homogeneous medium.

The shear sliding behaviour of adjacent elements, along the interfaces, being associated to a single degree of freedom, has been characterised according to a uniaxial nonlinear behavior, as clarified in the next section.

It is worth to notice that, the choice to concentrate the mechanical properties of the connected elements in zero-thickness interfaces is common to other discrete numerical approaches such for example the applied element method [30]-[34] and the rigid body spring model [35]-[38], however these latter strategies do not operate at the macro-scale.

Different levels of discretisation can be adopted in accordance to the chosen number of fibres along the masonry arch section. Furthermore, the presented procedure, for the evaluation of the interface stiffness matrix, can accommodate any nonlinear model chosen to represent the masonry constitutive law characterising the constitutive behaviour of each masonry fibre.

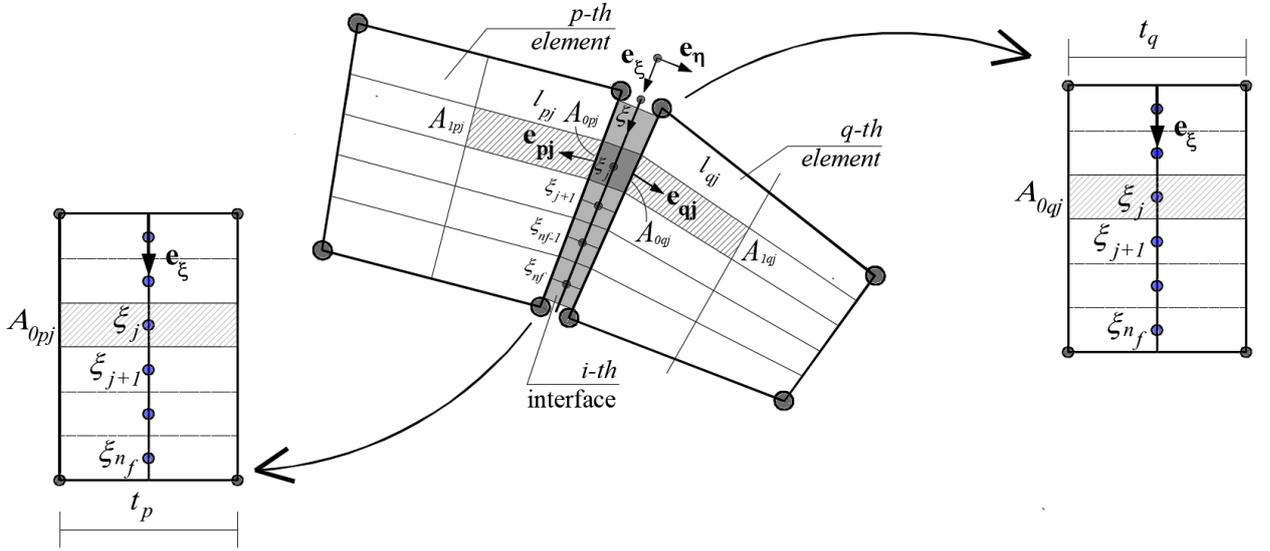

Figure 5. Fibre discretisation of the $i$-th interface and the adjacent macro-elements representation

### 3.1.1 The stiffness component orthogonal to the interface edge

The evaluation of the contribution of the $j$-th fibre, $j=1,\ldots,n_f$, to the tangent stiffness component $k_{T_v}(\xi)$ in the direction orthogonal to the interface of the $i$-th interface, is obtained as the combination in series of two contributions inherited by elements $p$ and $q$. For arches with linear non-uniform geometry and constant thickness of the element, the area of the $j$-th fibre varies linearly from $A_{0p_j}$ to $A_{1p_j}$ over the length $l_{p_j}$ on the $p$-th element and from $A_{0q_j}$ to $A_{1q_j}$ over the length $l_{q_j}$ on the $q$-th element, as indicated in Figure 5.

According to the fibre geometry the contributions $k_{p_j}, k_{q_j}$ to the tangent stiffness component of the $j$-th fibre relative to the $p$-th and $q$-th element are as follows:

$$k_{p_j} = \frac{E_{T p_j}}{\int_0^{l_{p_j}} \frac{dz}{A_{0 p_j} + \frac{z}{l_{p_j}}\left(A_{1 p_j} - A_{0 p_j}\right)}} \quad ; \quad k_{q_j} = \frac{E_{T q_j}}{\int_0^{l_{q_j}} \frac{dz}{A_{0 q_j} + \frac{z}{l_{q_j}}\left(A_{1 q_j} - A_{0 q_j}\right)}} \tag{14}$$

being $E_{T p_j}, E_{T q_j}$ the tangent modulus of the nonlinear constitutive behaviour of $j$-fibre relative to the $p$-th and $q$-th element, respectively. The integral appearing at the denominator of Eqs. (14), extended over the length of the fibre, by using a local coordinate $z$ along the fibre axis, can be evaluated leading to the following expressions:

$$k_{p_j} = \frac{E_{T p_j}\left(A_{1 p_j} - A_{0 p_j}\right)}{l_{p_j} \cdot \ln\left(\frac{A_{1 p_j}}{A_{0 p_j}}\right)} \quad \text{if } A_{1 p_j} \neq A_{0 p_j}; \quad k_{p_j} = \frac{E_{T p_j} A_{0 p_j}}{l_{p_j}} \quad \text{if } A_{1 p_j} \equiv A_{0 p_j}$$

$$k_{q_j} = \frac{E_{T q_j}\left(A_{1 q_j} - A_{0 q_j}\right)}{l_{q_j} \cdot \ln\left(\frac{A_{1 q_j}}{A_{0 q_j}}\right)} \quad \text{if } A_{1 q_j} \neq A_{0 q_j}; \quad k_{q_j} = \frac{E_{T q_j} A_{0 q_j}}{l_{q_j}} \quad \text{if } A_{1 q_j} \equiv A_{0 q_j} \tag{15}$$

In order to consider the actual orientation of the $i$-th interface, identified by the unit vector $\mathbf{e}_\eta$, with respect to the $j$-fibre relative to the $p$-th and $q$-th element, the two contributions to the tangent stiffness component, evaluated as in Eq. (15), are modified as follows:

$$k_{p_j}^i = k_{p_j} \left|\mathbf{e}_{p_j} \cdot \mathbf{e}_\eta\right| \quad ; \quad k_{q_j}^i = k_{q_j} \left|\mathbf{e}_{q_j} \cdot \mathbf{e}_\eta\right| \tag{16}$$

being $\mathbf{e}_{p_j}, \mathbf{e}_{q_j}$ the unit vectors identifying the $j$-th fibre orientation along the $p$-th and $q$-th element, as in Figure 5.

### 3.1.2 *The stiffness component along the interface edge*

In view of the kinematics of the macro-element, described in the previous section, the sliding mechanism of the two elements $p$, $q$ adjacent to the $i$-th interface is governed by a single relative displacement component $\hat{u}$, and the contribution of the $j$-th fibre, $j = 1, \ldots, n_f$, to the tangent stiffness component $k_{T_u}(\xi_j)$ along the $i$-th interface edge is here considered independent of the fibre position. Accordingly, the nonlinear behaviour of the interface ruling the sliding mechanism between the two edges of the interface is here calibrated by adopting an overall suitable nonlinear constitutive law of the cohesive-friction type.

Without loss of generality, only as a matter of example, for numerical application purposes a Mohr-Coulomb approach is followed by adopting a yielding dominium related to the actual contact area $A_c$ of the $i$-th interface, accounting for the presence of cracks, as follows:

$$F_y = c \cdot A_c + \mu \cdot N + g(\hat{u}^p) \tag{17}$$

where the mechanical parameters are the cohesion $c$ and the friction coefficient $\mu$, while $N = \sum_{j=1}^{n_f} f_v(\xi_j)$ is the resultant of the orthogonal forces $f_v$ at the $i$-th interface and $g(\hat{u}^p)$ represents an hardening function. For the case of rigid-plastic behaviour the tangent stiffness component $k_{T_u}(\xi_j)$ along the $i$-th interface is as follows:

$$\begin{aligned} k_{T_u}(\xi_j) &= \infty \quad \forall \; j = 1,\ldots,n_f \quad \text{for} \quad \sum_{j=1}^{n_f} f_u(\xi_j) < F_y \\ k_{T_u}(\xi_j) &= \frac{1}{n_f} \frac{dg(\hat{u}^p)}{d\hat{u}^p} \quad \forall \; j = 1,\ldots,n_f \quad \text{for} \quad \sum_{j=1}^{n_f} f_u(\xi_j) = F_y \end{aligned} \tag{18}$$

where $\sum_{j=1}^{n_f} f_u(\xi_j)$ is the resultant of the forces $f_u$ along the $i$-th interface direction.

It has to be noted that the adoption of the yielding dominium as in Eq. (17) somehow accounts for the influence of the internal forces normal to the $i$-th interface on the sliding mechanism. On the other hand, the influence of the sliding mechanism on the stiffness along the direction orthogonal to the interface is neglected. For the latter two reasons the out-of-diagonal terms of the tangent stiffness matrix in Eq. (10) are considered null. However, it has to be reminded that the mechanical model of the interface fibre discretization of the proposed macro-element can accommodate any bi-axial constitutive law able to account for the longitudinal-orthogonal mutual influence.

### *3.2. The macro-element in-plane shear diagonal stiffness*

The in-plane shear deformability of the proposed macro-element is controlled by a single Lagrangian parameter related to the angular distortion $\Gamma$ of the articulated quadrilateral, as introduced in sub-section 2.1 describing the element kinematics. The mechanical characterization of the shear element deformability is calibrated through a mechanical equivalence, introduced in this section, with reference to a geometrically consistent continuous plane model.

The calibration of the macro-element shear stiffness $K_\Gamma$ is performed by enforcing an equivalence with a homogeneous continuum model, i.e. an isotropic plate of the same geometry and subjected to a displacement field consistent to the kinematics of the articulated quadrilateral as in Figure 3b.

In order to enforce the equivalence between the discrete and the continuous model, the displacement field of the plate is first provided as function of the variation of angle $\Gamma$, ruling the deformation mode. It is worth to notice that the presented approach implies that the strain field consistent with a pure shear behavior is recovered in the case of regular quadrangular element.

The displacement field $\mathbf{u}^T(x,y) = \begin{bmatrix} u_x(x,y) & u_y(x,y) \end{bmatrix}$ of a generic point of the corresponding irregular plate is defined by its components along the $x$ and $y$ directions $u_x(x,y)$, $u_y(x,y)$ respectively, that can be expressed according to the intrinsic coordinates $\varsigma$ and $\lambda$, defined in the range $[-1,1]$, as follows:

$$u_x(\varsigma,\lambda) = \sum_{i=1}^{4} u_{ix} m_i(\varsigma,\lambda); \quad u_y(\varsigma,\lambda) = \sum_{i=1}^{4} u_{iy} m_i(\varsigma,\lambda) \tag{19}$$

being $u_{ix}, u_{iy}$, $i=1,\ldots 4$, the translational displacements of the nodes of the quadrilateral, Figure 3, and $m_i(\varsigma,\lambda)$ the classical bilinear polynomial functions given by:

$$m_1(\varsigma,\lambda) = \frac{(1-\varsigma)(1-\lambda)}{4}; \quad m_2(\varsigma,\lambda) = \frac{(1+\varsigma)(1-\lambda)}{4};$$
$$m_3(\varsigma,\lambda) = \frac{(1+\varsigma)(1+\lambda)}{4}; \quad m_4(\varsigma,\lambda) = \frac{(1-\varsigma)(1+\lambda)}{4} \tag{20}$$

Since the macro-element deformation does not depend on the rigid body motion it is sufficient to consider a kinematics in which one side of the quadrilateral is rigidly constrained. Without loss of generality, by constraining the first edge (between the nodes *1* and *2*), the displacement of vertices *3* and *4* only are considered, as a consequence the summations in Eq. (20) can be limited to the last two terms. The corresponding deformation field is given by

$$\varepsilon_x(\varsigma,\lambda) = \frac{\partial u_x(\varsigma,\lambda)}{\partial x}; \quad \varepsilon_y(\varsigma,\lambda) = \frac{\partial u_y(\varsigma,\lambda)}{\partial y}; \quad \gamma_{xy}(\varsigma,\lambda) = \frac{\partial u_y(\varsigma,\lambda)}{\partial x} + \frac{\partial u_x(\varsigma,\lambda)}{\partial y} \tag{21}$$

In view of Eqs. (20) and (21) the deformation field can be written as follows:

$$\varepsilon(\varsigma,\lambda) = \mathbf{B}(\varsigma,\lambda)\mathbf{u}_r \tag{22}$$

where

$$\varepsilon(\varsigma,\lambda) = \begin{bmatrix} \varepsilon_x(\varsigma,\lambda) \\ \varepsilon_y(\varsigma,\lambda) \\ \gamma_{xy}(\varsigma,\lambda) \end{bmatrix} \quad , \quad \mathbf{u}_r = \begin{bmatrix} u_{3x} \\ u_{3y} \\ u_{4x} \\ u_{4y} \end{bmatrix}$$

$$\mathbf{B}(\varsigma,\lambda) = \begin{bmatrix} \dfrac{\partial m_3(\varsigma,\lambda)}{\partial x} & 0 & \dfrac{\partial m_4(\varsigma,\lambda)}{\partial x} & 0 \\ 0 & \dfrac{\partial m_3(\varsigma,\lambda)}{\partial y} & 0 & \dfrac{\partial m_4(\varsigma,\lambda)}{\partial y} \\ \dfrac{\partial m_3(\varsigma,\lambda)}{\partial y} & \dfrac{\partial m_3(\varsigma,\lambda)}{\partial x} & \dfrac{\partial m_4(\varsigma,\lambda)}{\partial y} & \dfrac{\partial m_4(\varsigma,v)}{\partial x} \end{bmatrix} \tag{23}$$

According to the macro-element kinematics, described in section 2, the nodal displacement vector $\mathbf{u}_r$ can be expressed in terms of the variation of angle $\Gamma$ as follows:

$$\mathbf{u}_r = \mathbf{C}_r\,\Gamma \tag{24}$$

where the vector $\mathbf{C}_r$, in view of Eqs. (8), is given as:

$$\mathbf{C}_r = \begin{bmatrix} -\dfrac{l_4 \sin\alpha_4 \sin\alpha_2}{\sin\alpha_3} \\ -\dfrac{l_4 \sin\alpha_4 \cos\alpha_2}{\sin\alpha_3} \\ -l_4 \sin\alpha_1 \\ l_4 \cos\alpha_1 \end{bmatrix} \tag{25}$$

Accounting for Eqs. (24) and (25) the deformation field vector given by Eq.(22) can now be expressed as follows:

$$\varepsilon(\varsigma,\lambda) = \mathbf{B}(\varsigma,\lambda)\mathbf{C}_r\Gamma \tag{26}$$

Furthermore, by assuming a plane stress condition, the stress field $\boldsymbol{\sigma}^T(\varsigma,\lambda) = [\sigma_x(\varsigma,\lambda) \quad \sigma_y(\varsigma,\lambda) \quad \tau_{xy}(\varsigma,\lambda)]$, collecting normal $\sigma_x, \sigma_y$ and shear $\tau_{xy}$ stress components in the $x,y$ plane, related to a linear elastic isotropic constitutive law, is given by:

$$\boldsymbol{\sigma}(\varsigma,\lambda) = \mathbf{D}\,\varepsilon(\varsigma,\lambda) \tag{27}$$

where

$$\mathbf{D} = \frac{E}{1-\nu^2} \begin{bmatrix} 1 & -\nu & 0 \\ -\nu & 1 & 0 \\ 0 & 0 & 2(1+\nu) \end{bmatrix} \tag{28}$$

Once the strain and stress field have been obtained, the internal virtual work $\delta L_{int}$ can be written as follows:

$$\delta L_{int} = \int_{-1}^{1}\int_{-1}^{1} \boldsymbol{\sigma}^T(\varsigma,\lambda)\,\delta\boldsymbol{\varepsilon}(\varsigma,\lambda)\,J(\varsigma,\lambda)\,t\,d\varsigma\,d\lambda \tag{29}$$

where the Jacobian function $J(\varsigma,\lambda) = \frac{\partial x}{\partial \varsigma}\frac{\partial y}{\partial \lambda} - \frac{\partial x}{\partial \lambda}\frac{\partial y}{\partial \varsigma}$ has been introduced and $\delta(\cdot)$ indicates any virtual variation of the indicated quantity. Substitution of Eqs. (26) and (27) in Eq. (31) leads to:

$$\delta L_i = \Gamma \int_{-1}^{1}\int_{-1}^{1} \mathbf{C}_r^{\mathbf{T}}\mathbf{B}^{\mathbf{T}}(\varsigma,\lambda)\,\mathbf{D}\,\mathbf{B}(\varsigma,\lambda)\,\mathbf{C}_r\,t\,J(\varsigma,\lambda)\,d\varsigma\,d\lambda\,\delta\Gamma \tag{30}$$

being $t$ the constant thickness of the element. The double integral in Eq. (30) represents the scalar stiffness $K_\Gamma$ of a generic four-node plate with a constant thickness associated to the Lagrangian parameter $\Gamma$. The stiffness $K_\Gamma$ can be approximated by means of a Gaussian integration scheme in the space of the intrinsic coordinates $\varsigma$ and $\lambda$ of the plate leading to the following expression:

$$K_\Gamma = \sum_{k=1}^{N_G}\sum_{l=1}^{N_G} w_k\,w_l\,t\left[\mathbf{C}_r^{\mathbf{T}}\,\mathbf{B}^{\mathbf{T}}(\varsigma_k,\lambda_l)\,\mathbf{D}\,\mathbf{B}(\varsigma_k,\lambda_l)\,\mathbf{C}_r\right]J(\varsigma_k,\lambda_l) \tag{31}$$

where the coefficients $w_k, w_l$ represent the Gaussian weights and $N_G \times N_G$ is the number of Gaussian points adopted for the integration.

Once the elastic shear stiffness $K_\Gamma$ of the macro-element has been defined, based on the equivalent isotropic plate, its uniaxial nonlinear evolution is defined by suitable choices for the yielding domain able to account for the confinement action of adjacent elements. In particular, two possible yielding domains, suitable for masonry media, can be considered, namely the Mohr-Coulomb or the Turnsek and Cacovic [39] criteria. Further cyclic constitutive laws incorporating stiffness degradation can also be adopted, as reported in [23],[25], with reference to the rectangular macro-element.

## 4. Numerical applications

The proposed macro-element approach has been implemented in the software HiStrA [40], specifically devoted to nonlinear analyses of Historical Masonry Structures. The applications reported in the following aim at validating the proposed DMEM, both in the linear and nonlinear field, through a comparison with analytical, numerical and experimental results already available in the specific literature. In particular, the results of two different experimental campaigns have been taken into account.

The first application is relative to an experimental campaign, led by Ramos et al. on a masonry circular arch [41], for which both static and dynamic tests have been performed.

The second considered benchmark is relative to another circular masonry arch, extensively studied in the literature [42], by means of limit analysis and nonlinear finite element approaches.

The choice of the circular arches occurred to obtain a validation of the model with experimental and numerical data already available in the literature. However, the proposed approach is not limited to circular arches only. Each element has a generic quadrangular shape that can be adapted to different geometrical layouts according to an assumed mesh, similar to a FEM modelling.

Without loss of generality, the constitutive laws described in the previous sections are conveniently particularized as better described in the following.

The axial behaviour of each masonry fibre is characterised by an elastic-plastic behaviour with linear post peak softening branches whose ductility is governed by fracture energy values in tension, $G_t$, and compression, $G_m$, as qualitatively reported in Figure 6.

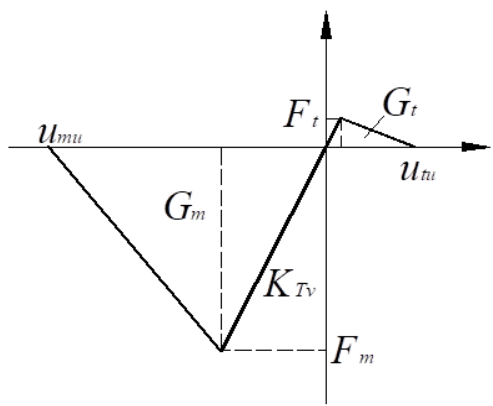

Figure 6. Constitutive law adopted for the axial behaviour of the fibers

The tensile $F_t$ and compressive $F_m$ strengths of each fibre are assumed to be related to the minimum cross area $A_{p_j \min}$ of the fibre along its span as a function of $f_{t_p}$ and $f_{m_p}$, respectively the tensile and compressive yielding strengths of the homogenised masonry medium of the *p*-th element. The compressive and tensile behaviour is also characterized by a linear softening associated to the corresponding fracture energies identified in Table 1 by the capital letter *G*. Consequently, the ultimate displacements $u_{t_u}(\xi_j), u_{m_u}(\xi_j)$ can be conveniently expressed as reported in Table 1.

Table 1. The mechanical characteristics of the nonlinear fibre equivalent to two adjacent fibres

| Elastic stiffness | Compressive and tensile yielding fibre forces | Compressive and tensile fibre ultimate displacements |
|---|---|---|
| $k_{T_v}(\xi_j) = \dfrac{k_{p_j}^i k_{q_j}^i}{k_{p_j}^i + k_{q_j}^i}$ | $F_t(\xi_j) = \min\left(A_{p_j \min} f_{t_p}, A_{q_j \min} f_{t_q}\right)$ $F_m(\xi_j) = \min\left(A_{p_j \min} f_{m_p}, A_{q_j \min} f_{m_q}\right)$ | $u_{t_u}(\xi_j) = \dfrac{G_{t_p} + G_{t_q}}{F_t(\xi_j)}$ $u_{m_u}(\xi_j) = \dfrac{G_{m_p} + G_{m_q}}{F_m(\xi_j)}$ |

With regard to the sliding behaviour a post-elastic linear softening is here employed, associated to a sliding fracture energy $G_s$. In particular, with reference to Eq. (18), the function $g(\hat{u}^p)$ is given by

$$g(\hat{u}^p) = \frac{(cA)^2}{2G_s} \hat{u}^p \tag{32}$$

Finally, for the diagonal shear mechanism, an elastic-perfectly plastic behaviour is considered.

### 4.1 *Simulation of an experimental campaign on a circular masonry arch subjected to static loads and dynamic characterization*

This experimental campaign has been conducted at the laboratory of the Civil Engineering Department of the University of Minho with the aim of identifying damage in masonry arches. To this purpose, a circular arch, built with clay bricks (100 x 50 x 25 mm³) bounded with mortar with poor mechanical properties [43], is considered. The geometrical layout of the investigated arch is reported in Figure 7, considering a width of the arch equal to 450 mm, however a detailed description of the specimen and the performed experiment can be found in [41]. In the experimental campaign modes and frequencies of vibrations have been identified in the undamaged configuration at first; then, in order to induce a damage in the arch, to be successively identified, several cycles of loading and unloading were performed by applying a concentrated vertical force located at the quarter of the span, as reported in Figure 7.

The first numerical simulations here performed aim at providing a model validation in the linear domain through a comparison in terms of eigen-properties. Aiming at performing both an experimental and numerical validation, the eigen-properties obtained by the proposed model have been compared with the experimental values as well as the results obtained through a plane linear FEM simulation performed by using the software environment SAP2000 [44].

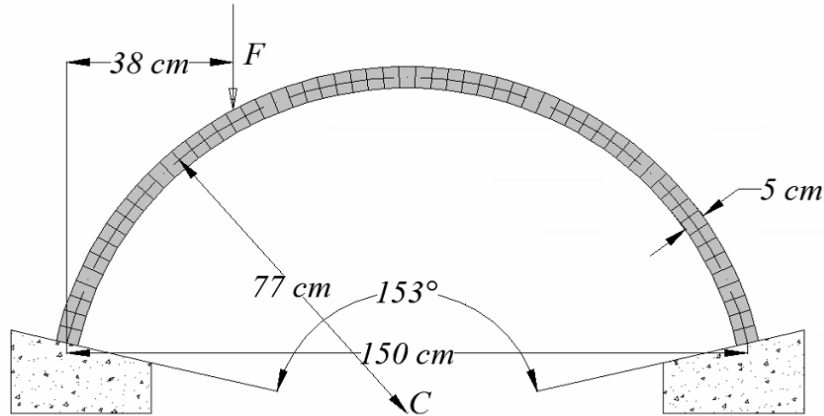

Figure 7. Geometric layout of the experimental test.

Table 2. Mechanical properties adopted for the numerical model of the arch

| $E$ [MPa] | $f_t$ [MPa] | $G_t$ [N/mm] | $f_m$ [MPa] | $G_m$ [N/mm] | $G$ [MPa] | $c$ [MPa] | $\mu$ | $G_s$ [N/mm] | $w$ [kN/m³] |
|---|---|---|---|---|---|---|---|---|---|
| 3790 | 0.25 | 0.02 | 7.8 | 90 | 1516 | 0.3 | 0.4 | 7.0 | 15 |

All the mechanical properties of the homogenized material, adopted in the numerical simulations according to the experimental data, are reported in Table 2. In particular $f_m$ and $f_t$ are the strengths in compression and tension, $G_m$ and $G_t$ are the compressive and tensile fractural energy values.

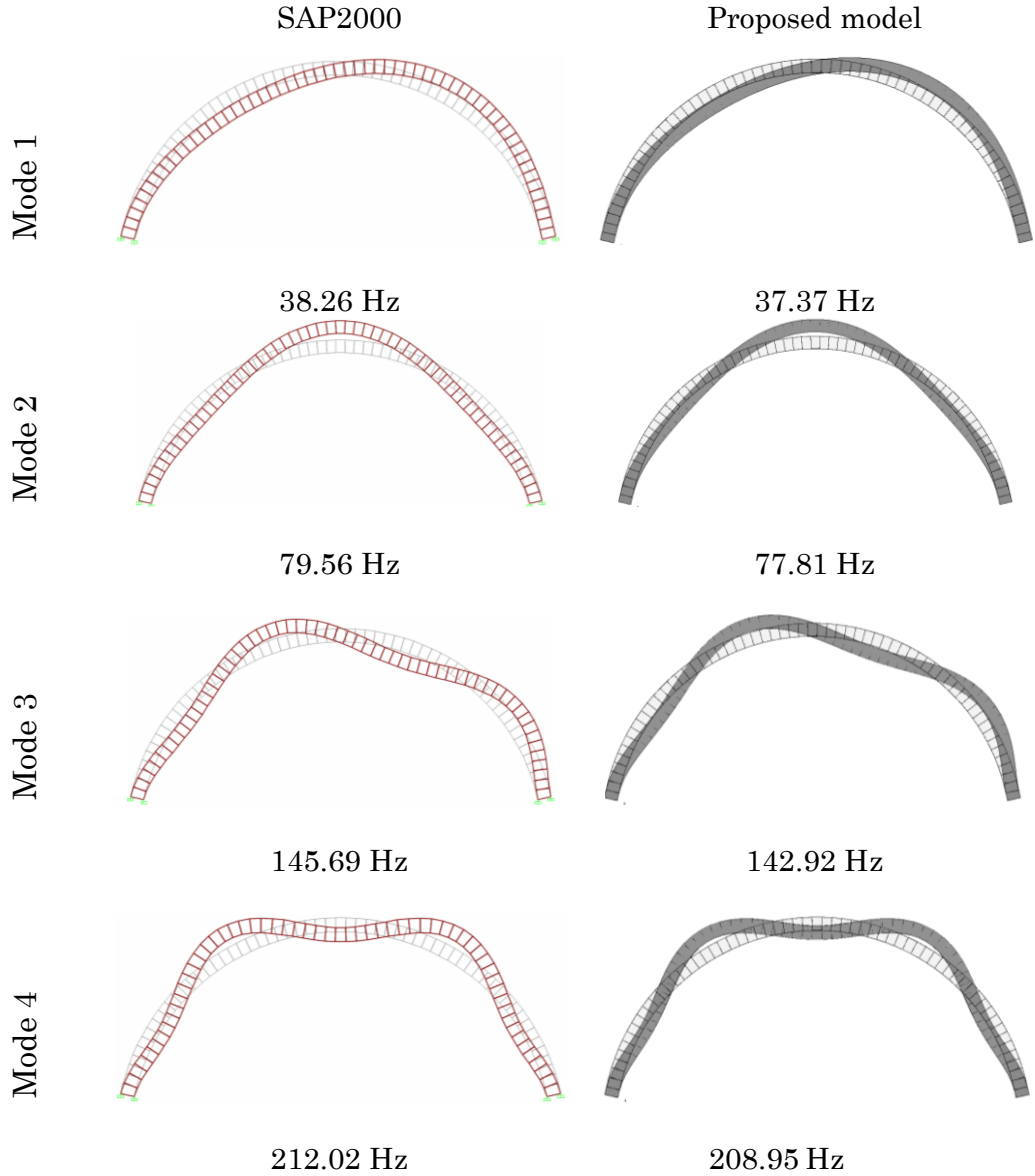

Figure 8. Comparisons of the first four modes of vibration obtained by the DMEM (HISTRA) and continuous FEM (SAP2000).

Finally the cohesion $c$, the friction coefficient $\mu$ and the fractural shear energy $G_s$ govern the Mohr-Coulomb yielding criterion and its ductility. The elastic properties, governed by the modulus $E$ and $G$, have been determined as suggested in [45].

The comparison in terms of the first four mode shapes is reported in Figure 8, where the corresponding natural frequencies are shown as well. It can be observed how the considered model is able to provide the same vibration modes obtained by the two-dimensional FEM analysis.

A more detailed comparison among the frequencies is reported in Table 3 in which the first four eigenvalues have been considered. The differences in terms of frequencies with respect to the experimental and the FEM results are within the limits of 7.90% and 2.32% respectively; it is worth to note that only the first three eigen-values have been experimentally evaluated.

Table 3. Comparison in terms of frequencies

| Modes | Experimental | | Continuous model (SAP2000) | | Proposed discrete model (HISTRA) | | | |
|---|---|---|---|---|---|---|---|---|
| | $T$ [s] | $f$ [Hz] | $T$ [s] | $f$ [Hz] | $T$ [s] | $f$ [Hz] | $Err_{exp}$ [%] | $Err_{SAP}$ [%] |
| 1 | 0.0281 | 35.59 | 0.0261 | 38.26 | 0.0268 | 37.37 | 5.00 | 2.32 |
| 2 | 0.0139 | 72.11 | 0.0126 | 79.56 | 0.0129 | 77.81 | 7.90 | 2.20 |
| 3 | 0.0071 | 140.08 | 0.0068 | 145.69 | 0.0070 | 142.92 | 2.03 | 1.90 |
| 4 | _ | _ | 0.0047 | 212.02 | 0.0048 | 208.95 | _ | 1.45 |

With the aim to evaluate the influence of the shear deformability of the macro-element in the linear range, a circular arch characterised by different aspect ratios $t/R$=0.1, 0.15, 0.20, 0.25 has been analysed considering the same radius and material properties of the previous example but different thicknesses. The eigen-values of the investigated arches have been than evaluated by accounting and/or ignoring the shear deformability. As reported in Table 4, frequency differences ranging from 1.56 % to 9.17 % are observed, highlighting the influence of the shear deformability, particularly for squat arches.

Table 4. The influence of the shear-diagonal deformability in the frequencies

| Modes | $t/R$=0.10 | | | $t/R$=0.15 | | | $t/R$=0.20 | | | $t/R$=0.25 | | |
|---|---|---|---|---|---|---|---|---|---|---|---|---|
| | $G$=1516 MPa | $G\to\infty$ | Diff [%] | $G$=1516 MPa | $G\to\infty$ | Diff [%] | $G$=1516 MPa | $G\to\infty$ | Diff [%] | $G$=1516 MPa | $G\to\infty$ | Diff [%] |
| 1 | 56.94 | 57.83 | 1.56 | 83.53 | 86.42 | 3.46 | 108.08 | 114.61 | 6.04 | 130.25 | 142.20 | 9.17 |
| 2 | 116.31 | 119.28 | 2.55 | 163.14 | 171.32 | 5.01 | 197.67 | 210.77 | 6.63 | 220.36 | 235.00 | 6.64 |
| 3 | 213.63 | 221.76 | 3.81 | 299.82 | 308.30 | 2.83 | 328.44 | 339.40 | 3.34 | 356.44 | 384.68 | 7.92 |
| 4 | 274.78 | 279.62 | 1.76 | 304.41 | 322.46 | 5.93 | 362.20 | 394.42 | 8.90 | 399.47 | 426.74 | 6.82 |

To validate the proposed approach in the nonlinear field, a nonlinear static analysis was performed on the arch. In order to simulate its actual behaviour, first the self weight was applied and then the concentrated load, according to the layout reported in Figure 7. The considered monitored displacement is the vertical component of the loaded point.

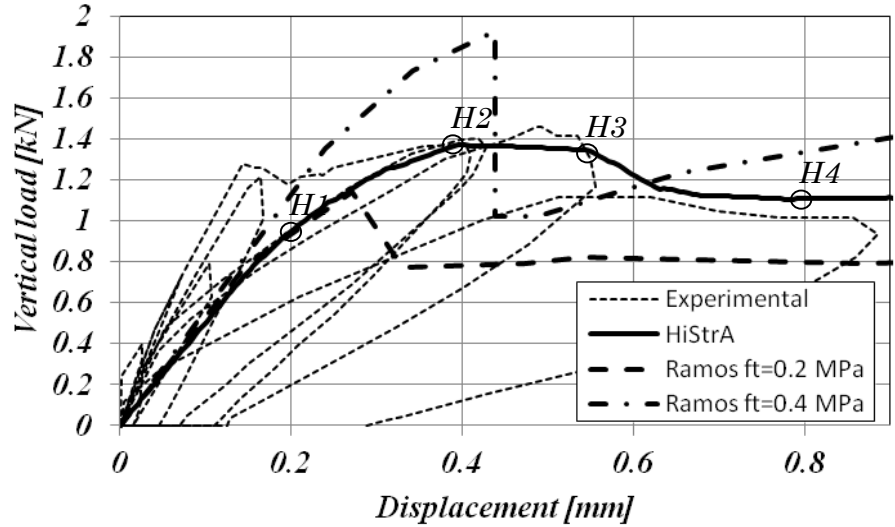

Figure 9. Capacity curves: comparison between the proposed model and the experimental results.

In Figure 9 the comparison in terms of load-displacement curves is reported, showing how the considered model provides a good prediction of the nonlinear behaviour of the experimental test, in terms of envelope curve. The peak load achieved during the cyclic experimental test is 1.45 kN, while in the monotonic numerical simulation performed with the proposed approach the peak load is 1.36 kN. The residual force at the end of the softening branch is about 1.05 kN against a value of about 1.01 kN in the experiment. In the same picture, the curves obtained by Ramos through a nonlinear FEM simulation [45], are reported. The FEM numerical simulations considered two different values of the tensile strength (in Figure 9 the cases of $f_t$=0.2 MPa and $f_t$=0.4 MPa are reported) and assumed a tensile fracture energy equal to 1/10 of the tensile strength (expressed in N/mm).

The nonlinear response of the first example is associated to the activation of four flexural hinges. According to the obtained numerical results all the hinges reach the limit tensile strength but do not attain to the ultimate compressive strength. As a result, the structure maintains a residual capacity that is related to the residual reaction of the plastic hinges. It is worth to notice that when the progressive reduction of the compressive area leads to the ultimate compressive strength, the residual capacity of the arch will necessary drop to zero.

In Figure 10 the numerical predictions are shown at different load levels corresponding to the sequential initial opening of the flexural hinge. A magnification factor of the deformed shapes equal to 100 is adopted. In the same Figure 10 the inelastic stored strain in the direction orthogonal to the interfaces, defined in [46], is reported in greyscale.

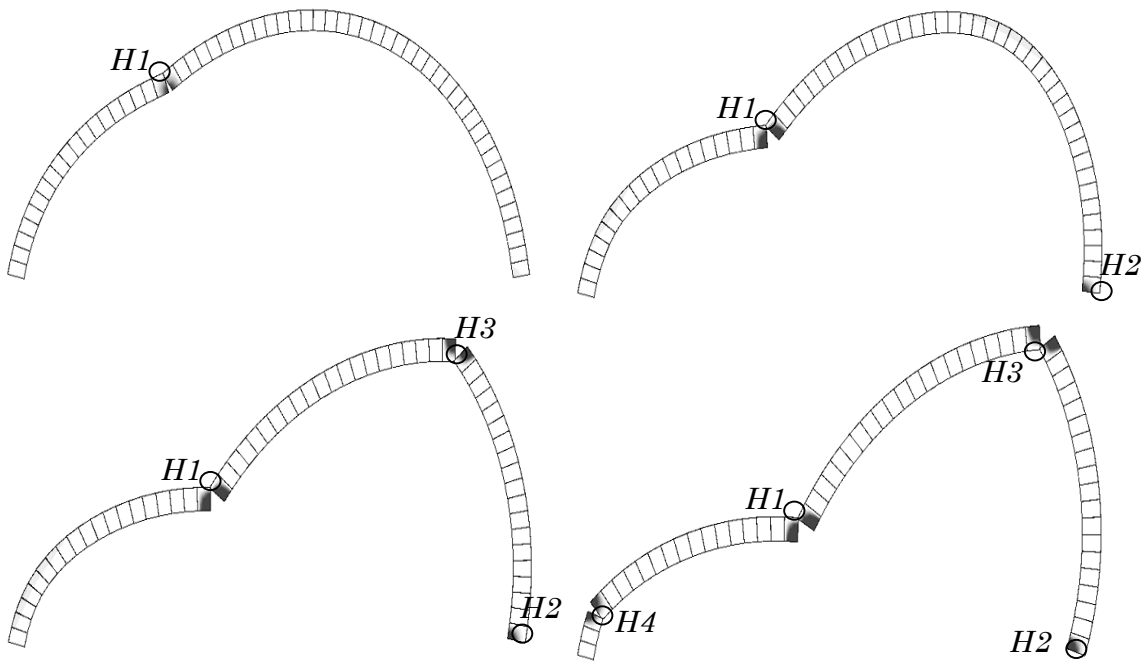

Figure 10.   Damage scenarios for different levels of the monitored displacement and final collapse mechanism .

Aiming at investigating the influence of the main parameters governing the nonlinear static behaviour of the arch, a sensitivity analysis with respect to the tensile strength and the tensile fracture energy has been reported in the following. Figure 11a reports the capacity curves obtained by considering four further values of the tensile strengths ($f_t$=0.1, 0.2, 0.3, 0.4 N/mm) together with the already performed analysis relative to $f_t$=0.25 MPa, and a tensile fracture energy $G_t$=0.02 N/mm. It can be observed how the tensile strength highly influences the ultimate load and the post-peak behaviour.

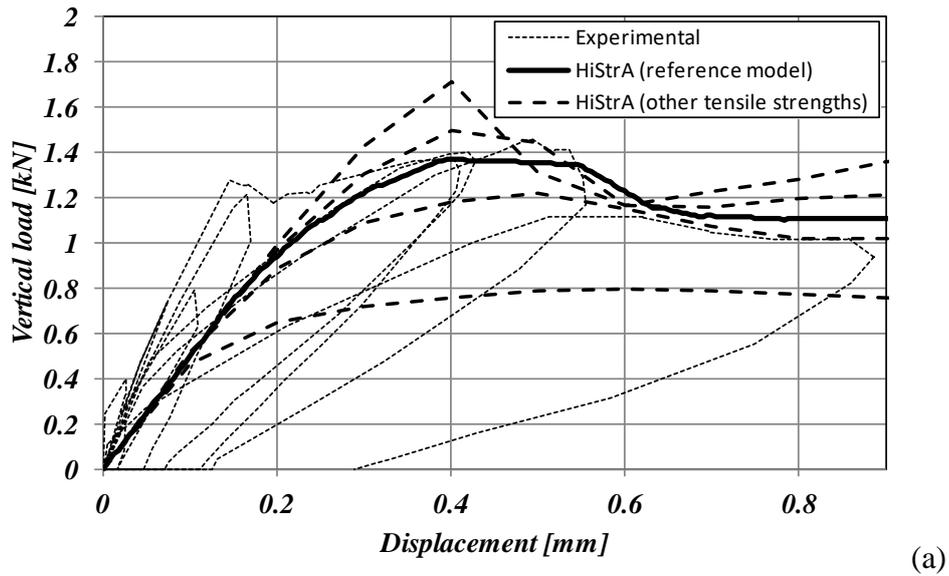

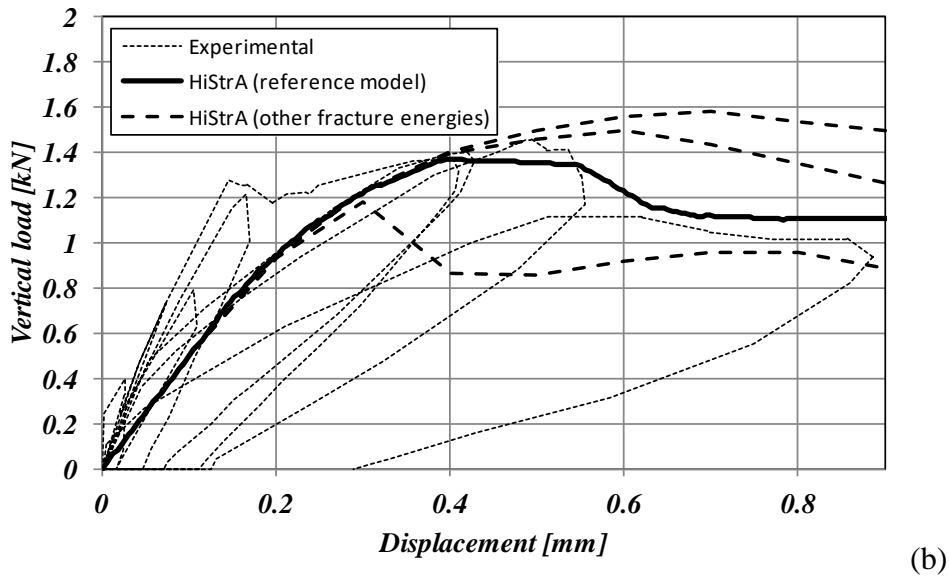

Figure 11. Capacity curves: comparison between the proposed model and the experimental results for different levels of (a) tensile strength and (b) tensile fracture energy.

The influence of different values of the fracture energy ($G_t$=0.01, 0.02, 0.03, 0.04 N/mm) for a fixed value of the peak tensile strength, $f_t$=0.25 MPa, is investigated in Figure 11b, it can be observed how both the peak load and the global ductility are strongly influenced by the assumed fracture energy.

## *4.2 Comparison with other numerical results on a benchmark circular masonry arch*

In this section a numerical validation of the proposed model is reported considering a set of elementary applications in order to compare the results of the proposed discrete model with the results of refined nonlinear FEM and limit analysis methods already published in literature. Namely, a stone circular arch bridge, which has been extensively studied in the past [42],[47],[48] is considered.

The main plane geometrical parameters, thickness, inner span and inner rise, are summarized Figure 12 and the width of the arch is equal to 1 m.

The finite element analysis model [42] consists of quadrilateral, four-node, bilinear, plane strain elements with two translational degrees of freedom per node. A typical value for the length of each finite element is 0.05 m. A total number of 3036 elements was used. In the FEM model unilateral interfaces are included between the parts of the structure and for the parametric investigation the number of uniformly distributed interfaces has been gradually increased. Large displacement effects are neglected and the arch is considered to be fixed to the ground.

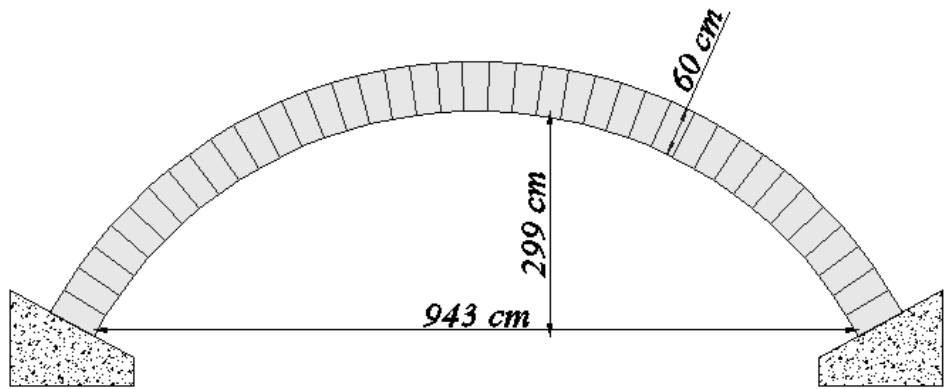

Figure 12.     Geometric layout of the experimental test.

The macro-element numerical model was implemented by considering a stone by stone discretization with 39 elements corresponding to 156 in-plane degrees of freedom. The mechanical properties of the stone blocks have been chosen according to those adopted in the FEM model reported in [42] and summarized in Table 5. The elastic deformation of the blocks is described by the Young modulus ($E$) and Poisson coefficient ($v$), the simplified hypotheses of zero tensile strength ($f_t$) and cohesion ($c$), and unlimited compressive strength ($f_m$), have been adopted for the bed joints. Furthermore unlimited ductility is considered for the sliding behaviour.

Table 5. Mechanical properties adopted for the numerical model of the arch

| Flexural behaviour (transversal N-Links) | | | | | Sliding behaviour (longitudinal N-Links) | |
|---|---|---|---|---|---|---|
| $E$ [MPa] | $v$ | $w$ [kN/m³] | $f_t$ [MPa] | $f_m$ [MPa] | $c$ [MPa] | $\mu$ |
| 5000 | 0.3 | 22 | 0 | ∞ | 0 | 0.6 |

The numerical simulations have been performed applying the self-weight of the arch at first, and then a concentrated vertical load with increasing amplitude. Two different load scenarios have been considered according to the position of the vertical load: in one case the load is applied at the mid-span (case 1), while in the second case at the quarter span (case 2). The results of the two analyses, in terms of collapse mechanisms, are reported in Figure 13. In both cases the behaviour is characterised by the occurrence of flexural hinges without the activation of sliding mechanisms along the stone interfaces. In the case of the mid-span applied load five plastic hinges occur (because of the symmetry of the geometry of the arch and of the load), while four plastic hinges occur in the other case. The collapse mechanisms are consistent with those obtained in the simulations already reported in the literature [42].

The force-displacement results are reported in Figure 14 in terms of total vertical base reaction ($F$) vs the vertical displacement of the application point of the external force.

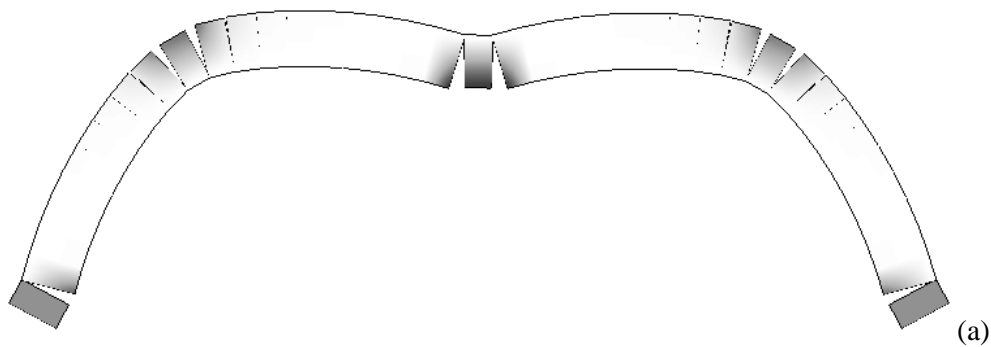

(a)

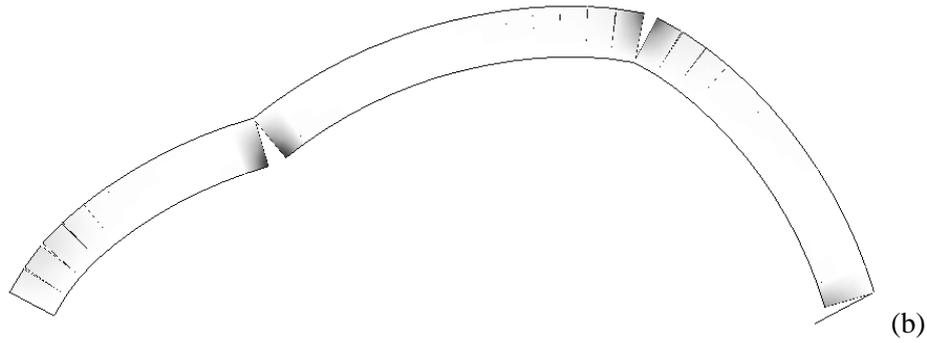

Figure 13.    Collapse mechanisms of the arch: (a) mid-span load and (b) quarter of span load.

A good agreement with the limit and FEM analyses in terms of ultimate load can be recognised, however the spread plasticity FEM approach shows a different trend of the pushover curves with respect to the proposed DMEM.

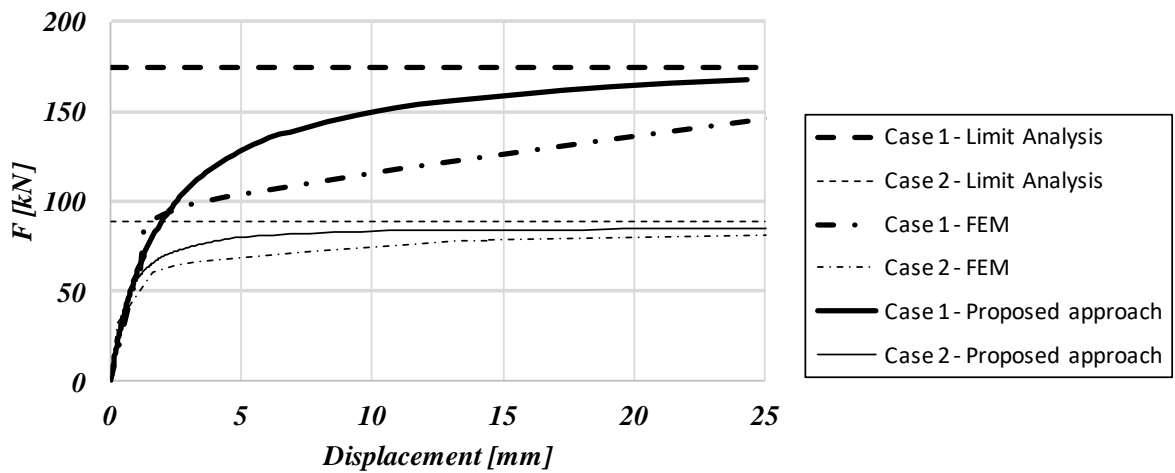

Figure 14.    Pushover curves of the arch and comparisons of the proposed model (continuous lines) with limit analysis (dashed lines) and FEM approach (dash dot lines): mid-span load (thick lines) and quarter of span load (thin lines).

The collapse mechanism of the arch can be dominated by the flexural or the shear behaviour according to the value of the friction coefficient attributed to the interfaces (and keeping the other mechanical properties according to Table 5). In order to identify the influence of the value of the friction coefficient on the limit load, as well as on the corresponding failure scenarios, several numerical analyses have been performed gradually reducing its value for both the load scenarios.

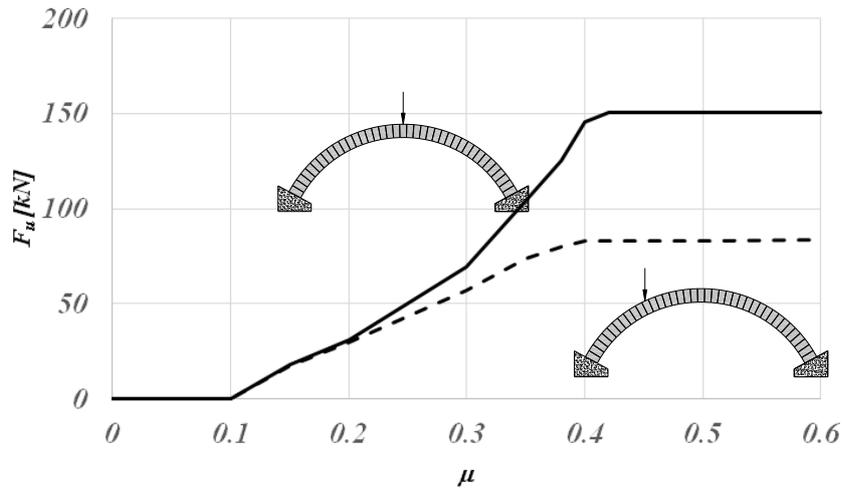

Figure 15.    Ultimate load vs friction coefficient: mid-span load
(continuous line) and quarter of span load (dashed line).

In Figure 15 the collapse value of the applied load is shown as a function of the friction coefficient. The collapse value of the applied load has been conventionally referred to the load that produces a vertical displacement of 10 mm in the section where the load is applied. It can be observed that for values of the friction coefficient greater than about $\mu_c=0.4$, the load is constant and equal to 150 $kN$, in this range only flexural hinges occur, being the friction coefficient sufficient to prevent shear-sliding along the interfaces. When the friction coefficient is lower than $\mu_c$ the collapse mechanism involves a shear sliding close to the section in which the load is applied and the value of the load progressively reduces, following a roughly linear trend, until the value of 0.1 corresponding to a case in which neither the self-weight loading condition can be accomplished. The bifurcation value of friction coefficient $\mu_c$ is in the range 0.38-0.4 for the case of applied eccentric load and 0.4-0.42 for the case of mid-span load.

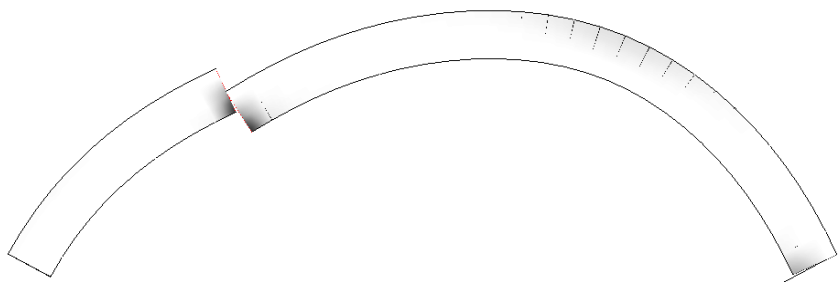

Figure 16.    Collapse mechanisms of the arch with the occurring of sliding:
quarter of span load and $\mu=0.3$.

In Figure 16, a typical collapse mechanism of the arch associated to the occurring of sliding is reported. This is characterized by two plastic hinges involving the shear and the flexural behaviours located in correspondence of the applied load and at the right base; in addition a spread flexural damage can be observed at the extrados symmetrically with respect to the applied concentrated load.

A further parameter which strongly influences the ultimate carrying capacity of the structure is the load position (*x*), particularly in the case of masonry arch bridges. In the application reported in the following the role of the position of the vertical load in the ultimate capacity of the arch is investigated. Two values of the friction coefficient have been considered, namely $\mu$=0.6 and $\mu$=0.3.

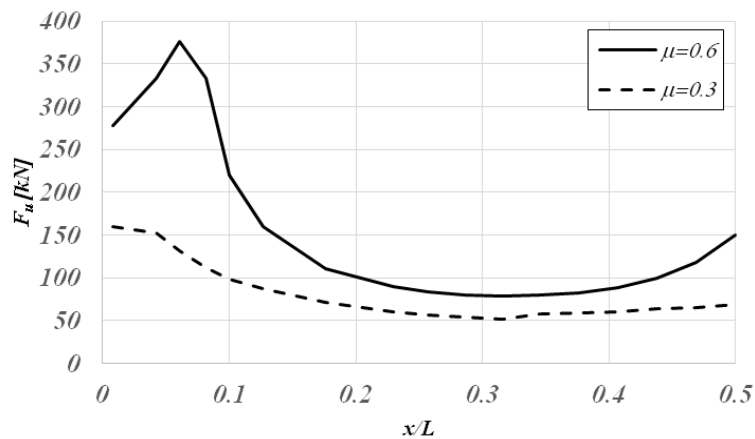

Figure 17.    Ultimate load vs load position for two different values of the friction coefficients: $\mu$ =0.6 (continuous line) and $\mu$=0.3 (dashed line).

The obtained results are summarized in Figure 17, where the ultimate load (related to an ultimate conventional displacement equal to 25 mm) is reported as a function of the normalized abscissa of the load position considering one half of the arch. The obtained results show that the minimum ultimate load is related to positions of the concentrated vertical load close to the value of the normalised abscissa 0.3 for both the investigated cases.

The collapse mechanisms associated to some significant positions of the load are reported in Table 6, with the indication of the corresponding limit value. The cases corresponding to the higher friction coefficient show collapse mechanisms characterised by the occurrence of four flexural hinges (with the exception of the case of central load characterised by five hinges). The cases corresponding to the lower friction coefficient are all characterised by flexural or shear hinges depending on the position of the load.

Table 6. Influence of the load position: collapse mechanisms

| $x/L$ | $\mu=0.3$ | $\mu=0.6$ |
|---|---|---|
| 0.1 | 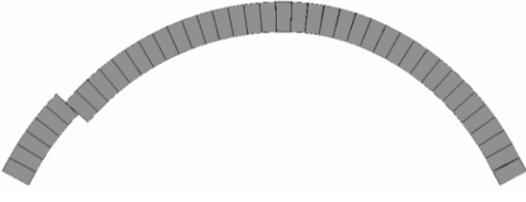 $F_u$=98.42 kN | 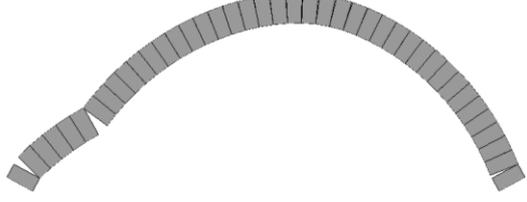 $F_u$=219.99 kN |
| 0.25 | 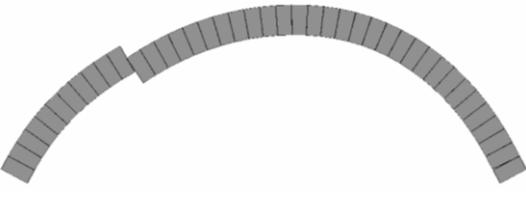 $F_u$=56.90 kN | 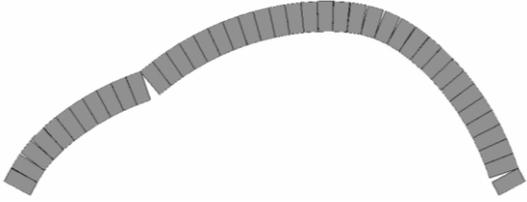 $F_u$=83.46 kN |
| 0.4 | 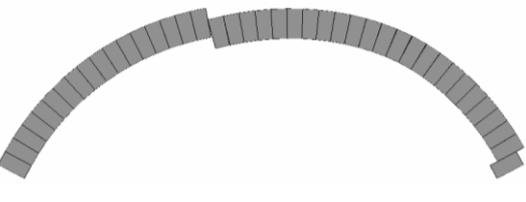 $F_u$=59.82 kN | 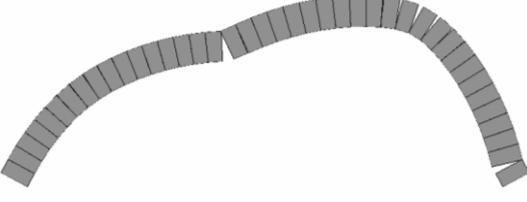 $F_u$=82.58 kN |
| 0.5 | 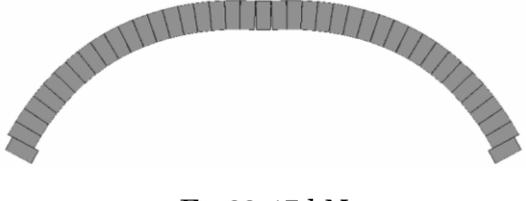 $F_u$=69.47 kN | 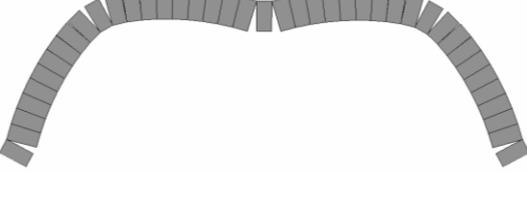 $F_u$=150.52 kN |

The arch behaviour has been also investigated considering different values of tensile strength and fracture energy. The first parametric investigation considers a fixed value of tensile strength, $f_t = 0.02\,\text{MPa}$, and different fracture energy values between the limit cases of $G_t \to 0$ (brittle behaviour) and $G_t \to \infty$ (infinite ductility), for the mid-span (Figure 18a) and one-quarter span positions (Figure 18b) of the load. The influence of the tensile fracture energy has been assessed by considering four different values ( $G_t = 0.01, 0.05, 0.1, \infty \, \text{N/mm}$). It can be observed how the fracture energy affects the ultimate load of the arch although maintaining a good ductility behaviour also for a low value of fracture energy, in this latter case the ultimate load is close to the lower bound provided by the limit analysis, Figure 14.

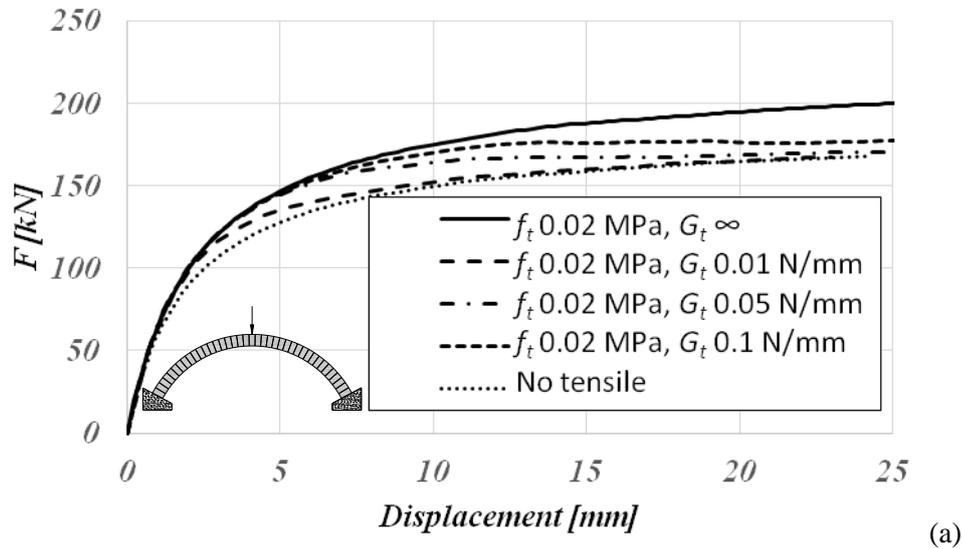

(a)

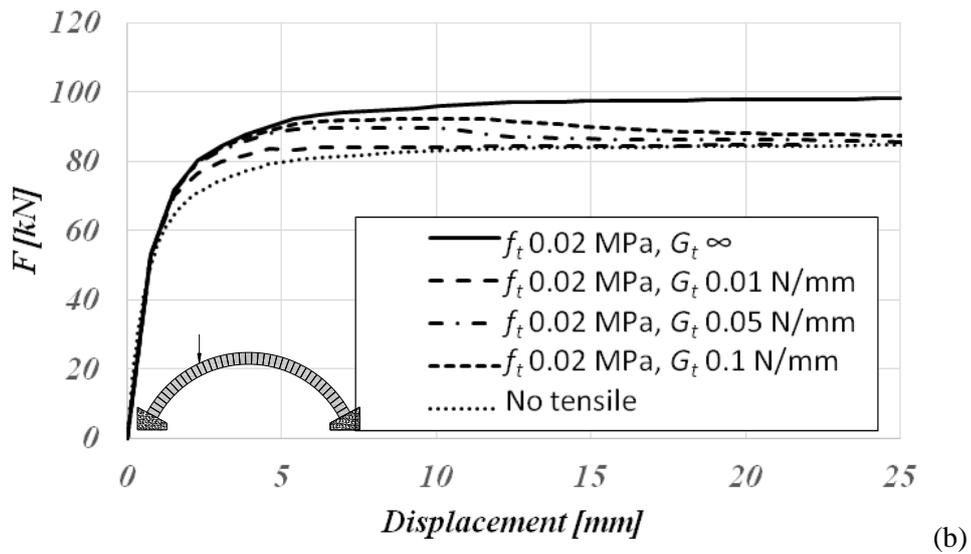

(b)

Figure 18. Influence of the tensile fracture energy: (a) mid-span load and (b) quarter of span load.

The influence of the tensile strength has been investigated by considering the load scenarios for the values of the tensile strength $f_t = 0, 0.01, 0.02, 0.04 \, \text{MPa}$ and perfectly ductile behaviour $G_t \rightarrow \infty$, the results are reported in Figure 19. It can be observed how the tensile strength strongly influences the ultimate load capacity of the arch in both the load configurations.

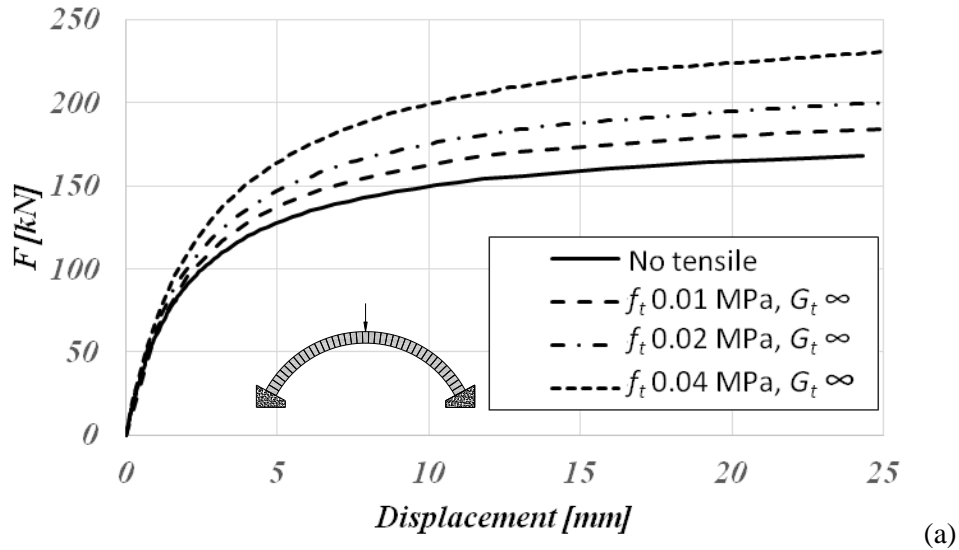

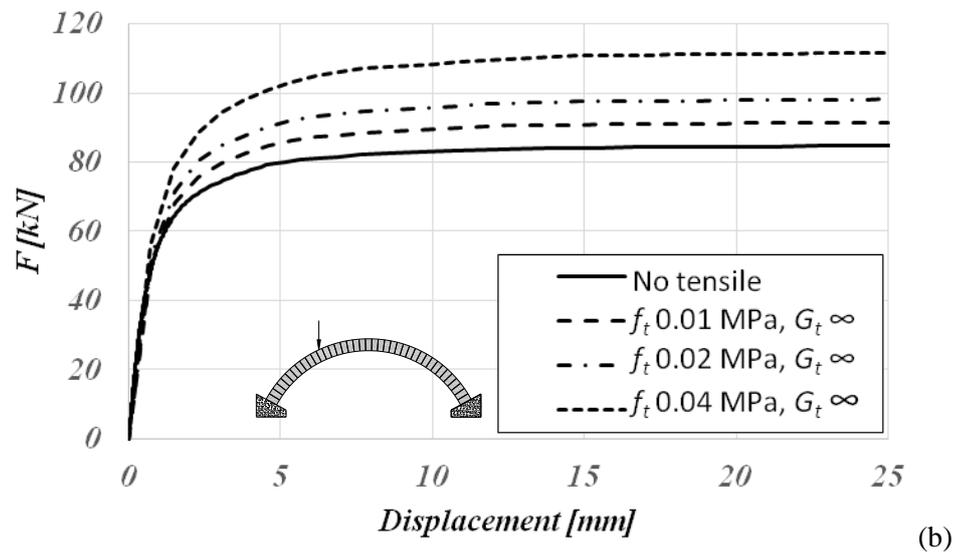

Figure 19. Influence of the tensile strength: (a) mid-span load and (b) quarter of span load.

## 5. Conclusions

In this paper a Discrete Macro-Element (DME) approach for the assessment of the nonlinear behaviour of masonry arches is presented. The method can be regarded as a discrete method in which each element possesses an internal deformability and represents the corresponding masonry element, at the macro-scale, according to a simplified kinematics. A single degree of freedom accounts for the internal shear deformability, three further degrees of freedom describe the rigid motion of each macro-element. Namely, each macro-element can be assimilated to a hinged quadrilateral with an internal deformability and contouring interfaces which govern the interaction with the adjacent elements. The calibration of the model requires few parameters in order to define the basic masonry mechanical properties. The equivalence between the macro-masonry arch portion and the macro-element is based on a very simple fibre calibration strategy that makes the interpretation of the numerical results straightforward and unambiguous.

In spite of its simplicity and ability to limit the needed computational effort to perform numerical simulations, this approach appears to be able to simulate the main in-plane failure mechanisms of masonry arch structures, also in presence of irregular geometry layouts.

Several comparisons with experimental and benchmark numerical results demonstrate the reliability and suitability of the model in the evaluation of the bearing capacity of arched masonry structures.

It is worth to mention that geometric nonlinearity could significantly affect the results, particularly with reference to the post-peak behaviour in the nonlinear range. The adopted linear kinematics represents a limit of the current formulation, the extension to a general macro-element strategy, accounting for material and geometrical nonlinearities in curved geometry masonry structures, is currently a work in progress.

## 6. Acknowledgement

This research has been supported by the Italian Network of SeismicEngineering University Laboratories (ReLUIS). This work is part of theNational Research Project"Advanced mechanical modelling of newmaterials and structures for the solution of 2020 Horizon challenges"(2017–2020), supported by MIUR, Grant No. 2015JW9NJT, Scientificcoordinator, prof. M. Di Paola, prot. n. 2015JW9NJT_017.

## 7. References

[1] S. Huerta, Galileo was wrong: the geometrical design of masonry arches, Nexus Netw J 8(2) (2006) 25–6.
[2] A.S. Gago, J. Alfaiate, A. Lamas, The effect of the infill in arched structures: Analytical and numerical modelling, Engineering Structures, 33 (5) (2011) 1450-1458.

[3] J. Heyman, The stone skeleton: structural engineering of masonry architecture, Cambridge University Press 1995.

[4] J. Heyman, On the rubber vaults of the middle ages, and other matters, Gaz Beaux-Arts 71 (1968)177–88.

[5] J. Heyman, The safety of masonry arches, International Journal of Mechanical Sciences, 11 (4) (1969) 363-382,IN3-IN4,383-385.

[6] J. Heyman, The estimation of the strength of masonry arches, Proc Inst Civ Engrs 1980;69:921.

[7] J. Heyman, The masonry arch, Ellis Horwood: Chichester; 1982.

[8] I. Caliò, A. Greco, D. D'Urso, Free vibrations of spatial Timoshenko arches, Journal of Sound and Vibration, Volume 333, Issue 19, 2014, Pages 4543-4561.

[9] I. Caliò, ,A. Greco, D. D'Urso, Structural models for the evaluation of eigen-properties in damaged spatial arches: a critical appraisal, Arch Appl Mech (2016) 86: 1853.

[10] F. Cannizzaro, A. Greco,S. Caddemi,I. Caliò, Closed form solutions of a multi-cracked circular arch under static loads, International Journal of Solids and Structures (2017), 121, pp. 191-200.

[11] I. Caliò, D. D'Urso, A. Greco, The influence of damage on the eigen-properties of Timoshenko spatial arches, Computers and Structures, (2017) 190, pp. 13-24.

[12] A. Cavicchi, L. Gambarotta, Collapse analysis of masonry bridges taking into account arch-fill interaction, Engineering Structures 27 (4) (2005) 605-615.

[13] A. Thavalingam, N. Bicanic, J.I. Robinson, D.A. Ponniah, Computational framework for discontinuous modelling of masonry arch bridges, Computers and Structures, 79 (19) (2001) 1821-1830.

[14] J.V. Lemos, Discrete element modelling of the seismic behaviour of stone masonry arches, In: Pande GN, Middleton J, Kralj B, editors. Computer methods in structural masonry — 4. London: E & FN Spon; 1998. p. 220–7.

[15] A.R. Tóth, Z. Orbán, K. Bagi, Discrete element analysis of a stone masonry arch, Mechanics Research Communications, 36 (4) (2009) 469-480.

[16] V. Sarhosis, D.V. Oliveira, J.V. Lemos, P.B. Lourenco, The effect of skew angle on the mechanical behaviour of masonry arches, Mechanics Research Communications, 61 (2014) 53-59.

[17] E. Rizzi, F. Rusconi, G. Cocchetti, Analytical and numerical DDA analysis on the collapse mode of circular masonry arches, Engineering Structures, 60 (2014) 241-257.

[18] R. Dimitri, F. Tornabene, A parametric investigation of the seismic capacity for masonry arches and portals of different shapes, Engineering Failure Analysis (2015) 1-34.

[19]  R. Dimitri, L. De Lorenzis, G. Zavarise, Numerical study on the dynamic behavior of masonry columns and arches on buttresses with the discrete element method, Engineering Structures, 33 (2011) 3172-3188.

[20]  L. De Lorenzis, R. Dimitri, J. Ochsendorf, Structural study of masonry buttresses: the stepped form, ICE Proceedings – Structures and Buildings, 165(9) (2012) 499-521.

[21]  L. De Lorenzis, R. Dimitri, J. Ochsendorf, Structural study of masonry buttresses: the trapezoidal form, ICE Proceedings – Structures and Buildings, 165(9) (2012) 483-498.

[22]  Y. Zhang, L. Macorini, B.A. Izzuddin, Mesoscale partitioned analysis of brick-masonry arches, Engineering Structures, 124 (2016) 142-166, ISSN 0141-0296, http://dx.doi.org/10.1016/j.engstruct.2016.05.046.

[23]  I. Caliò, M. Marletta, B. Pantò, A new discrete element model for the evaluation of the seismic behaviour of unreinforced masonry buildings, Engineering Structures 40 (2012) 327-338.

[24]  Caddemi S., Caliò I., Cannizzaro F., Pantò B., A new computational strategy for the seismic assessment of infilled frame structures, Proceedings of the 14th International Conference on Civil, Structural and Environmental Engineering Computing, CC 2013, (2013) Civil-Comp Proceedings.

[25]  I. Caliò, B. Pantò, A macro-element modelling approach of Infilled Frame Structures, Computers and Structures 143 (2014) 91–107.

[26]  B. Pantò, F. Cannizzaro, S. Caddemi, I. Caliò, 3D macro-element modelling approach for seismic assessment of historical masonry churches, Advances in Engineering Software 97 (2016) 40–59.

[27]  A. Zucchini, P.B. Lourenço, A micro-mechanical model for the homogenisation of masonry International Journal of Solids and Structures 39 (12) (2002) 3233-3255.

[28]  C. Wu, H. Hao, Derivation of 3D masonry properties using numerical homogenization technique, International Journal for Numerical Methods in Engineering, 66 (11) (2006) 1717-1737.

[29]  A. Bacigalupo, L. Gambarotta, Computational two-scale homogenization of periodic masonry: Characteristic lengths and dispersive waves, Computer Methods in Applied Mechanics and Engineering 213-216 (2012) 16-28.

[30]  K. Meguro, H. Tagel-Din, Applied element method for structural analysis: Theory and application for linear materials Structural Engineering/Earthquake Engineering 17 (1) (2000) 21-35.

[31]  K. Meguro, H. Tagel-Din, Applied Element Method for Simulation of Nonlinear Materials: Theory and Application for RC Structures, Structural Engineering/Earthquake Engineering 17 (2) (2000) 137-148.

[32] K. Meguro, H. Tagel-Din, Applied element simulation of RC structures under cyclic loading, Journal of Structural Engineering 127 (11) (2001) 1295-1305.

[33] P. Mayorka, K. Meguro, Modeling Masonry Structures using the Applied Element Method, Seisan Kenkyu. Japan: Institute of Industrial Science, The University of Tokyo 55 (6) (2003) 123–126. ISSN 1881-2058.

[34] A. Furukawa, J. Kiyono, K. Toki, Proposal of a numerical simulation method for elastic, failure and collapse behaviors of structures and its application to seismic response analysis of masonry walls, Journal of Disaster Research 6 (1) (2011) 51-69.

[35] F. Nagashima, I. Fumihito, Application of RBSM to slipping problem of friction-type joints, Memoirs of Faculty of Technology, Tokyo Metropolitan University (33) (1983) 3317-3327.

[36] S. Casolo, Modelling the out-of-plane seismic behaviour of masonry walls by rigid elements, Earthquake Engineering and Structural Dynamics 29 (12) (2000) 1797-1813.

[37] S. Casolo, Modelling in-plane micro-structure of masonry walls by rigid elements, International Journal of Solids and Structures 41 (13) (2004) 3625-3641.

[38] K.M. Dolatshahi, A.J. Aref, Two-dimensional computational framework of meso-scale rigid and line interface elements for masonry structures, Engineering Structures 33 (12) (2011) 3657-3667.

[39] V. Turnsek, F. Cacovic, Some experimental result on the strength of brick masonry walls. In: Proceedings of the 2nd International Brick Masonry Conference; Stoke-on-Trent (1971) 149–56.

[40] HiStrA (Historical Structure Analysis) software. HISTRA s.r.l, Catania, Italy. Release 17.2.3, April 2015. http://www.grupposismica.it

[41] L.F. Ramos, G. De Roeck, P.B. Lourenço, A. Campos-Costac, Damage identification on arched masonry structures using ambient and random impact vibrations Engineering Strctures 32 (2010) 146-162.

[42] G.A. Drosopoulos, G.E. Stavroulakis, C.V. Massalas, Limit analysis of a single span masonry bridge with unilateral frictional contact interfaces, Engineering Structures 28 (2006) 1864-1873.

[43] I. Basilio, Strengthening of arched masonry structures with composite materials. Ph.D. Thesis. Portugal: University of Minho. Available from: www.civil.uminho.pt/masonry; 2007

[44] CSI Analysis Reference Manual for SAP2000, Computers and Structures Inc., 2007.

[45] L.F. Ramos, Damage identification on masonry structures based on vibration signatures. Ph.D. Thesis. Portugal: University of Minho. Available from: www.civil.uminho.pt/masonry; 2007.

[46]   B. Pantò, F. Cannizzaro, S. Caddemi, I. Caliò, Numerical and experimental validation of a 3D macro-model for the in-plane and out-of-plane behavior of unreinforced masonry walls, International Journal of Architectural Heritage 11(7) (2017) 946–964.

[47]   M. Betti, G.A. Drosopoulos, G.E. Stavroulakis, On the collapse analysis of single span masonry/stone arch bridges with fill interaction. In: Proceedings of the 5th international conference on arch bridges ARCH'07. 2007. p. 617–24.

[48]   A. Cavicchi, L. Gambarotta, Load carrying capacity of masonry bridges: numerical evaluation of the influence of fill and spandrels. In: Proceedings of the 5th international conference on arch bridges ARCH'07. 2007. p. 609–16.